\documentclass[10pt,a4paper,twocolumn]{article}

\usepackage[T1]{fontenc}
\usepackage[utf8]{inputenc}
\usepackage[english]{babel}
\usepackage{lmodern}
\usepackage{microtype}

\usepackage[
a4paper,
left=18mm,
right=18mm,
top=20mm,
bottom=22mm
]{geometry}

\usepackage{amsmath}
\usepackage{amssymb}
\usepackage{amsfonts}
\usepackage{steinmetz}

\usepackage{graphicx}
\usepackage{array}
\usepackage{tabularx}
\usepackage{booktabs}

\graphicspath{{./}{./image/}}

\usepackage{xcolor}
\usepackage{listings}
\usepackage{lstautogobble}
\usepackage{tcolorbox}
\tcbuselibrary{skins,listings,xparse,breakable,hooks,theorems,fitting,raster}

\definecolor{SLcodebg}{HTML}{F6F6F6}
\definecolor{SLcodeborder}{HTML}{DEDEDE}
\definecolor{SLcodekeyword}{HTML}{2F5597}
\definecolor{SLcodestring}{HTML}{A33A3A}
\definecolor{SLcodecomment}{HTML}{5F7F5F}

\newcommand{\SLbodyfont}{%
	\fontsize{7.5pt}{8.5pt}\selectfont\ttfamily%
}

\lstdefinestyle{Rstyle}{
	language=R,
	basicstyle=\SLbodyfont,
	keywordstyle=\color{SLcodekeyword},
	stringstyle=\color{SLcodestring},
	commentstyle=\color{SLcodecomment}\upshape,
	numbers=none,
	numberstyle=\scriptsize\color{gray},
	numbersep=6pt,
	showstringspaces=false,
	breaklines=true,
	columns=fullflexible,
	keepspaces=true,
	tabsize=2,
	aboveskip=0pt,
	belowskip=0pt
}

\lstdefinestyle{Routput}{
	basicstyle=\SLbodyfont,
	numbers=none,
	showstringspaces=false,
	breaklines=true,
	columns=fullflexible,
	keepspaces=true,
	aboveskip=0pt,
	belowskip=0pt
}

\tcbset{
	SL/base/.style={
		enhanced,
		frame hidden,
		boxrule=0pt,
		colback=SLcodebg,
		arc=0mm,
		outer arc=0mm
	},
	SL/output/.style={
		SL/base,
		left=2mm,right=2mm,
		top=0.6mm,bottom=0.6mm
	},
	SL/floating-title/.style={
		fonttitle=\bfseries,
		coltitle=black,
		colbacktitle=white,
		titlerule=0pt
	}
}

\newtcblisting[auto counter,number within=section]{rcode}[2][]{
	SL/base,
	listing only,
	listing options={style=Rstyle,xleftmargin=0pt},
	breakable,
	colback=SLcodebg,
	colbacktitle=white,
	colframe=SLcodebg,
	coltitle=black,
	frame hidden,
	boxrule=0pt,
	arc=0mm,
	outer arc=0mm,
	left=5mm,right=2mm,
	top=0.4mm,bottom=0.4mm,
	title={\textbf{BOX~\thetcbcounter}\quad|\quad #2},
	fonttitle=\normalfont\fontsize{9}{11.5}\selectfont,
	titlerule=0pt,
	toptitle=0.7mm,bottomtitle=0.7mm,
	lefttitle=2mm,righttitle=2mm,
	#1
}

\newcommand{\SLoutputdescription}{}
\newcommand{\routputcaption}[2]{%
	\gdef\SLoutputdescription{#2}%
}

\tcbset{
	SL/result/.style={
		enhanced,
		breakable,
		frame hidden,
		colback=SLcodebg,
		colbacktitle=white,
		coltitle=black,
		boxrule=0pt,
		arc=0mm,
		outer arc=0mm,
		titlerule=0pt,
		fontupper=\SLbodyfont,
		fonttitle=\normalfont\fontsize{9}{11.5}\selectfont\raggedright,
		toptitle=0.6mm,bottomtitle=0.6mm,
		lefttitle=2mm,righttitle=2mm,
		left=2mm,right=2mm,
		top=0.6mm,bottom=0.6mm
	},
	SL/result-numbered/.style={
		SL/result,
		title={\textbf{OUTPUT~\thetcbcounter}\quad|\quad\SLoutputdescription}
	},
	SL/result-plain/.style={
		SL/result,
		title={\textbf{OUTPUT}\quad|\quad\SLoutputdescription}
	}
}

\newtcblisting{routput}{
	SL/result-plain,
	listing only,
	listing options={style=Routput}
}

\newtcolorbox{routputbox}{
	SL/result-plain
}

\newtcblisting[auto counter]{routputref}[1][]{
	SL/result-numbered,
	listing only,
	listing options={style=Routput},
	#1
}

\newtcolorbox[use counter from=routputref]{routputboxref}[1][]{
	SL/result-numbered,
	#1
}

\lstdefinestyle{SLcheat}{
	language=R,
	basicstyle=\SLbodyfont,
	numbers=none,
	autogobble=true,
	xleftmargin=0pt,
	resetmargins=true,
	tabsize=2,
	breaklines=true,
	breakatwhitespace=false,
	columns=fullflexible,
	keepspaces=true,
	showstringspaces=false,
	aboveskip=0pt,
	belowskip=0pt
}

\usepackage{colortbl}
\usepackage[hidelinks]{hyperref}

\begin{document}
	
	\twocolumn[{%
		\begin{minipage}{\textwidth}
			
			{\centering
				{\LARGE\bfseries
					Super Learner: A Practical Tutorial\par}
				
				\vspace{0.9em}
				
				{\normalsize
					\href{https://orcid.org/0009-0004-8380-3198}
					{Eliezer Pita Zambrano}\textsuperscript{1}
					\quad
					\href{https://orcid.org/0000-0001-6683-5164}
					{Miguel Angel Luque-Fernandez}\textsuperscript{1,2}
					\par}
				
				\vspace{0.7em}
				
				{\small
					\textsuperscript{1}
					Department of Statistics and Operations Research,
					University of Granada, Granada, Spain\par
					
					\textsuperscript{2}
					Inequalities in Cancer Outcomes Network,
					London School of Hygiene and Tropical Medicine,
					London, UK\par}
				
				\vspace{0.4em}
				
			}
			
			\vspace{1.2em}
			
			\noindent\textbf{Abstract}\par
			\vspace{0.35em}
			
			\noindent\small
			The selection of an optimal predictive model is a pervasive
			challenge in biostatistics and epidemiology. Relying on a single
			\textit{a priori} algorithm risks overfitting and reduced
			generalizability when applied to new settings. Super Learner is a
			flexible ensemble machine-learning method that mitigates these
			risks by combining multiple candidate algorithms into a single
			prediction model. Using cross-validation, it defines a weighted
			combination of learners that minimizes a specified loss function,
			ensuring performance asymptotically equal to or better than the
			best candidate algorithm. This tutorial provides a comprehensive,
			practical guide to implementing Super Learner in R for applied
			researchers. Using a simulated epidemiological dataset with a
			binary outcome, we demonstrate the step-by-step construction of
			the ensemble. We cover the foundational statistical concepts,
			basic implementation, evaluation through cross-validation,
			hyperparameter tuning, feature screening, and performance
			optimization for metrics such as the Area Under the ROC Curve
			(AUC). Additionally, we detail advanced usage, including
			computational parallelization and the integration of custom
			algorithms and screening procedures. The guide also includes a
			consolidated section on best practices for specifying SL, covering
			loss function selection, effective sample size determination,
			CV scheme definition, library curation, and discrete vs. ensemble
			SL decisions, plus a practical reference to the \texttt{caret}
			package for model preprocessing and comparison. This guide equips
			epidemiologists and clinical researchers with the tools to
			confidently incorporate robust machine learning techniques into
			their predictive modeling workflows.
			
			\par\vspace{0.55em}
			
			
			\par\vspace{0.5em}
			
			\noindent\textbf{Code availability:}
			The complete R code to reproduce all examples in this tutorial
			is available at
			\url{https://github.com/migariane/SL_Tutorial}.
			
		\end{minipage}
		
		\vspace{1.4em}
	}]
	
\section{Introduction}\label{sec1}

A common challenge in the use of predictive models is that their performance may deteriorate when they are applied outside the conditions, populations, or settings in which they were developed and evaluated \cite{Collinse074819}. This behavior may arise from multiple factors, including shifts in the data distribution or population characteristics that were not adequately represented during model training \cite{Zhang2022Shifting}. In addition, the relative performance of learning methods may vary according to the characteristics and structure of the data; therefore, an algorithm that performs well under certain conditions may not necessarily retain that advantage in other settings \cite{ALI2006119,Uddin2024}. This dependence makes the \textit{a priori} selection of a single method difficult and may lead to increasingly complex model specifications in an attempt to adapt the model to the available data, thereby increasing the risk of overfitting and capturing sample-specific variation rather than generalizable patterns \cite{Pavlou2016}. \\
In clinical settings, the appropriate selection and evaluation of predictive models is particularly important because their predictions may contribute to clinical decision-making. An overfitted model may exhibit apparently strong performance in the data used for its development but show reduced predictive ability when applied to new individuals, resulting in overly optimistic estimates of its performance \cite{Riley2020,Collinse074819}. \\
Super Learner (SL) is an ensemble learning algorithm that combines predictions from multiple candidate algorithms with the aim of improving predictive performance. Using a library of models with different structures and degrees of flexibility, SL relies on cross-validation to evaluate their performance and estimate a weighted combination that minimizes a specified loss function. This procedure allows the predictive strengths of different algorithms to be leveraged without relying on the \textit{a priori} selection of a single model \cite{vanDerLaan2007,Wolpert1992,Breiman1996}.\\
Phillips et al.\cite{Phillips2023} provided practical guidelines for specifying a well-performing SL, including the selection of candidate learners, cross-validation schemes, performance metrics, and meta-learners. However, translating these general guidelines into workflow-ready clinical and epidemiological research requires bridging theoretical specifications with practical execution. This tutorial complements existing guidance by providing a step-by-step framework tailored specifically for applied epidemiologists, biostatisticians, and clinical researchers. By progressively connecting foundational concepts to code implementation, model evaluation, and result interpretation, we equip researchers with actionable tools to make, justify, and execute key methodological decisions in real-world analytical pipelines. \\
This tutorial provides a comprehensive, practical guide to implementing Super Learner in R for applied research. It progresses from foundational concepts through advanced applications, covering: (1) introduction to prediction challenges and the Super Learner framework; (2) SL fundamentals and statistical foundations; (3) practical implementation with simulated epidemiological data; (4) basic SL fitting; (5) predictions and performance evaluation; (6) cross-validated SL assessment; (7) hyperparameter tuning; (8) feature screening for high-dimensional data; (9) performance optimization for AUC and related metrics; (10) computational parallelization; (11) custom learner and screener integration; (12) full epidemiological application; (13) application to continuous outcomes; (14) dose-response estimation with continuous outcomes; (15) comparison with traditional methods (\texttt{caret} \cite{Kuhn2008Caret}); (16) consolidated best practices for specifying SL in epidemiological studies, integrating Phillips et al.'s \cite{Phillips2023} five key decisions (loss function, effective sample size, CV scheme, diverse library, discrete vs. ensemble SL) with library curation, CV scheme selection, and bridging to causal inference with TMLE; (17) quick-reference summary tables; and (18) SL quick reference cheat sheet. Each section bridges theoretical concepts with R code implementation, equipping researchers with actionable tools for real-world predictive modeling workflows. \\
The complete R code to reproduce all examples and figures in this tutorial is available at \url{https://github.com/migariane/SL_Tutorial}.
 
\section{SL Fundamentals}\label{sec2}
SL is an ensemble machine-learning method that uses $V$-fold cross-validation (CV) to estimate the performance of a set of candidate algorithms. Based on these estimates, it determines a weighted combination of their predictions by optimizing a loss function to obtain a predictor whose performance is asymptotically at least as good as that of the best candidate algorithm in the library \cite{vanDerLaan2007,vanDerLaanDudoit2006}.
\subsection{Statistical Foundations}
In a clinical study, a vector of covariates ($\mathbf{W}_i$) and an outcome ($Y_i$) are observed for each $i$. The observed data for each individual can be represented as
\begin{align}\label{eq1}
	O_i = (\mathbf{W}_i, Y_i), \qquad i = 1, \ldots, n.
\end{align}
The observations are assumed to arise from the same data-generating process, represented by an unknown probability distribution $P_0$, such that

\begin{align}\label{eq2}
	O_1,\ldots,O_n \overset{\mathrm{i.i.d.}}{\sim} P_0.
\end{align}

The prediction problem consists of estimating a function $\psi$ that maps covariate values to a prediction of the outcome:

\begin{align}\label{eq3}
	\psi: \mathcal{W} \rightarrow \mathbb{R}.
\end{align}

Here, $\psi(\mathbf{w})$ denotes the prediction corresponding to the covariate values $\mathbf{W}=\mathbf{w}$.
Prediction rules are compared using a loss function $L$. The risk of a prediction rule is defined as its expected loss:

\begin{align}\label{eq4}
	R(\psi) = E_{P_0}\left[L\left(Y,\psi(\mathbf{W})\right)\right].
\end{align}

The expectation is taken with respect to the true data-generating distribution $P_0$.
The optimal prediction rule, with respect to the specified loss function, is defined as the rule that minimizes the expected risk:

\begin{align}\label{eq5}
	\psi_0 = \arg\min_{\psi} R(\psi).
\end{align}

\subsection{Estimation Procedure}
SL starts from a collection of candidate learning algorithms.\cite{NaimiBalzer2018} Let $\mathcal{A}_k$, for $k=1,\ldots,K$, denote a learning algorithm that, when trained on a data set, produces a fitted prediction rule $\widehat{\psi}_k$. The collection of candidate algorithms defines the library

\begin{align}\label{eq6}
	\mathcal{L} = \left\{\mathcal{A}_1,\ldots,\mathcal{A}_K\right\}.
\end{align}

Because the true distribution $P_0$ is unknown, the population risk $R(\psi)$ cannot be evaluated directly. SL therefore uses CV to estimate the predictive performance of the candidate algorithms and determine their combination within the ensemble.\cite{vanDerLaan2007} \\
To implement $V$-fold CV, the $n$ observations are divided into $V$ subsets. For each iteration $v=1,\ldots,V$, let $V(v)$ denote the validation set and $T(v)$ its complementary training set. Each candidate algorithm $\mathcal{A}_k$ is fitted using $T(v)$ and generates predictions for the observations contained in $V(v)$. Because these predictions are obtained from models that were not fitted using the corresponding validation observations, they constitute cross-validated, or \textit{out-of-fold}, predictions. The predictions obtained from the $K$ candidate algorithms are collected in an $n \times K$ matrix $\mathbf{Z}$,

\begin{align}\label{eq7}
	\mathbf{Z} = (\mathbf{Z}_1,\ldots,\mathbf{Z}_K),
\end{align}

where $\mathbf{Z}_k$ contains the cross-validated predictions obtained from algorithm $\mathcal{A}_k$.
Given the matrix $\mathbf{Z}$, SL combines the candidate algorithms' predictions using a set of weights. Let $\alpha=(\alpha_1,\ldots,\alpha_K)$ denote the vector of weights assigned to the $K$ candidate algorithms. These weights are estimated through a meta-learning procedure, referred to as the \textit{meta-learner}, which seeks the combination that minimizes the cross-validated risk:

\begin{align}\label{eq8}
	\widehat{\alpha}
	=
	\arg\min_{\alpha} R_{CV}(\alpha).
\end{align}

Here, $R_{CV}(\alpha)$ denotes the risk associated with the weighted combination defined by $\alpha$. The specific form of this criterion depends on the meta-learning procedure used to estimate the ensemble weights. For the standard convex combination, the weights satisfy

\begin{align}\label{eq9}
	\alpha_k \geq 0,
	\qquad
	\sum_{k=1}^{K} \alpha_k = 1.
\end{align}

Once the weights have been estimated, each candidate algorithm is refitted using the full sample. For a new set of covariate values $\mathbf{w}$, the final SL prediction is defined as

\begin{align}\label{eq10}
	\widehat{\psi}_{SL}(\mathbf{w})
	=
	\sum_{k=1}^{K}
	\widehat{\alpha}_k
	\widehat{\psi}_k(\mathbf{w}).
\end{align}

\begin{rcode}[label={code:weighted-procedure}]{Implementing the weighted SL procedure}
	# Simulate cross-validated predictions Z (n x K matrix)
	set.seed(123, kind = "Mersenne-Twister")
	n <- 400
	K <- 3
	Z <- matrix(runif(n * K), nrow = n, ncol = K)
	colnames(Z) <- c("SL.glmnet", "SL.randomForest", "SL.mean")
	# Outcome vector
	Y <- rbinom(n, 1, 0.44)
	# Meta-learner: Non-negative least squares (NNLS) for convex combination
	library(nnls)
	fit_nnls <- nnls(Z, Y)
	alpha <- fit_nnls$x / sum(fit_nnls$x)  # normalize to sum to 1
	names(alpha) <- colnames(Z)
	print(round(alpha, 3))
	# Predict Candidate Learners on New Data (Test Set)
	n_new <- 50 # Number of new test observations to predict
	# Generate individual candidate predictions for new test instances
	# (In real workflows, replace these with predict(fit_model, newdata = X_new))
	psi_glmnet <- runif(n_new) 
	psi_rf     <- runif(n_new)
	psi_mean   <- rep(mean(Y), n_new)
	# Combine candidate predictions into a matrix matching Z's column structure
	Z_new <- cbind(
	SL.glmnet       = psi_glmnet,
	SL.randomForest = psi_rf,
	SL.mean         = psi_mean
	)
	# Final SL Ensemble Predictions
	# Option A: Explicit Manual Indexing 
	# Multiplies each algorithm's prediction vector by its assigned weight by name
	psi_SL_manual <- (alpha["SL.glmnet"]       * psi_glmnet) +
	(alpha["SL.randomForest"] * psi_rf)     +
	(alpha["SL.mean"]         * psi_mean)
	# Option B: Matrix Multiplication (Vectorized / Production) ---
	# Computes the exact same linear combination across all rows simultaneously
	psi_SL_matrix <- drop(Z_new %*% alpha)
	# Output Results and Verify Equivalence
	# Inspect first few predictions from both methods
	results_comparison <- head(data.frame(
	Manual_Indexing = psi_SL_manual,
	Matrix_Product  = psi_SL_matrix
	))
	cat("--- First Few SL Predictions ---\n")
	print(results_comparison)
	cat("\n")
\end{rcode}

\routputcaption{box:nnls-output}{Estimated ensemble weights from the NNLS meta-learner on simulated cross-validated predictions.}

\begin{routputboxref}[label={box:nnls-output}]
	\ttfamily
	\centering
	\begin{tabular}{lr}
		& weight \\
		SL.glmnet       & 0.307 \\
		SL.randomForest & 0.498 \\
		SL.mean         & 0.195 \\
	\end{tabular}
\end{routputboxref}

Box~\ref{code:weighted-procedure} illustrates the core weighted SL procedure. The matrix $\mathbf{Z}$ contains the cross-validated (out-of-fold) predictions from each of the $K$ candidate algorithms. The meta-learner uses Non-Negative Least Squares (NNLS) to solve for the weight vector $\boldsymbol{\alpha}$ that minimizes the squared error loss

\begin{align}\label{eq11}
	\widehat{\boldsymbol{\alpha}} = \arg\min_{\boldsymbol{\alpha} \geq \mathbf{0}} \| Y - \mathbf{Z} \boldsymbol{\alpha} \|_2^2,
\end{align}

subject to the non-negativity constraint $\alpha_k \geq 0$. The solution is then normalized so that $\sum_{k=1}^K \alpha_k = 1$, producing a convex combination. This ensures that the ensemble weights are interpretable as relative contributions and that the final prediction $\widehat{\psi}_{SL}$ is a weighted average of the candidate predictions. The \texttt{nnls} package implements the Lawson-Hanson algorithm \cite{LawsonHanson1995} for this constrained optimization problem. Once the weights are estimated, each candidate algorithm is refitted on the full training data, and the final SL prediction for new data is obtained by taking the weighted sum of the individual predictions.

\subsection{Meta-Learner Comparison: NNLS vs. Other Methods}\label{sec:meta-learner-comparison}
The \texttt{SuperLearner} package provides several built-in meta-learners (methods) for estimating the ensemble weights, each corresponding to a different loss function and optimization approach.\cite{Phillips2023} Table~\ref{tab:meta-learners} summarizes the main options.
\begin{center}
	\begin{table*}[!t]
		\centering
		\caption{Built-in meta-learners in the \texttt{SuperLearner} package.}
		\label{tab:meta-learners}
		\begin{tabularx}{\textwidth}{l l l X}
			\hline
			Method & Loss Function & Constraints & Typical Use Case \\
			\hline
			\texttt{method.NNLS}      & Squared Error (L2) & $\alpha_k \geq 0$, $\sum \alpha_k = 1$ & Default; continuous outcomes, general purpose \\
			\texttt{method.NNLS2}     & Squared Error (L2) & $\alpha_k \geq 0$, $\sum \alpha_k = 1$ & Same as NNLS but enforces sum-to-1 \textit{during} optimization \\
			\texttt{method.CC\_LS}     & Squared Error (L2) & $\alpha_k \geq 0$, $\sum \alpha_k = 1$ & Convex combination via quadratic programming (identical to NNLS2) \\
			\texttt{method.NNloglik}   & Negative Log-Likelihood & $\alpha_k \geq 0$, $\sum \alpha_k = 1$ & \textbf{Binary outcomes}; uses binomial likelihood \\
			\texttt{method.CC\_nloglik} & Negative Log-Likelihood & $\alpha_k \geq 0$, $\sum \alpha_k = 1$ & Binary outcomes; alternative implementation \\
			\texttt{method.AUC}        & $-$AUC (maximize AUC) & $\alpha_k \geq 0$, $\sum \alpha_k = 1$ & \textbf{Binary outcomes}; directly optimizes discrimination \\
			\hline
		\end{tabularx}
	\end{table*}
\end{center}
NNLS (\texttt{method.NNLS}) is the default meta-learner because the Lawson-Hanson algorithm solves the constrained least-squares problem efficiently with no tuning parameters, guarantees a global minimum for the convex problem $\min_{\alpha \geq 0} \|Y - Z\alpha\|_2^2$ (normalization preserves the optimum), yields interpretable convex-combination weights, and performs robustly for both continuous and binary outcomes. Use \texttt{method.NNLS2} or \texttt{method.CC\_LS} when the sum-to-1 constraint must be enforced during optimization (quadratic programming via \texttt{quadprog}); \texttt{method.NNloglik} or \texttt{method.CC\_nloglik} for binary outcomes when calibrated probabilities are the goal (minimizing binomial deviance, the proper scoring rule); and \texttt{method.AUC} when discrimination/ranking is the sole objective (maximizes cross-validated AUC, requires \texttt{nloptr}). Practical recommendation: start with NNLS for a quick baseline; compare with NNloglik for risk prediction needing calibrated probabilities; compare with AUC for pure discrimination; finally, use \texttt{CV.SuperLearner()} to externally evaluate and select the best meta-learner for your data.

\section{SL practical guide using small simulated data}\label{sec:sl-guide-smalldata}
In R, SL is available through the \texttt{SuperLearner} package \cite{vanDerLaan2007}. A more recent implementation, the \texttt{sl3} package \cite{Coyle2021sl3}, is part of the \texttt{tlverse} ecosystem and provides a modern, object-oriented interface to Super Learning built on the R6 class system. While \texttt{sl3} offers advantages such as machine learning pipelines, delayed computation, and tighter integration with the targeted learning framework, this tutorial uses the classical \texttt{SuperLearner} package for several reasons: it is available on CRAN (whereas \texttt{sl3} must be installed from GitHub), it has a simpler function-based API that is more accessible to new users, it has a larger established user base with extensive online resources and examples, and its procedural interface requires less familiarity with object-oriented programming in R.
To reproduce the examples presented throughout this tutorial, additional packages are required alongside the core package. These packages provide some of the candidate algorithms, tools for performance evaluation, and auxiliary functions  used throughout the analysis.

\begin{rcode}[label={code:install}]{Installing required packages}
	# Core package
	install.packages("SuperLearner")
	# Packages used throughout this tutorial
	install.packages(c(
	"glmnet", "randomForest", "ranger", "xgboost", "earth", "gam",
	"gbm", "rpart", "nnet", "e1071", "ipred", "ada",
	"ggplot2", "ROCR", "pROC", "caret", "data.table", "mlbench",
	"RhpcBLASctl", "rmutil", "vcd", "nnls", "Matrix", "tmle"
	))
\end{rcode}

The first command (Box~\ref{code:install}) installs the core \texttt{SuperLearner} package, whereas the second installs additional R packages required for subsequent examples. The specific role of each package is described when it is introduced. Once installed, the \texttt{SuperLearner} package must be loaded into the current R session (Box~\ref{code:load-sl}).

\begin{rcode}[label={code:load-sl}]{Loading the SuperLearner package}
	library(SuperLearner)
\end{rcode}

\subsection{Simulated Data}
This tutorial is developed using a small simulated data set. The data represent an epidemiological setting with a binary outcome, such as the presence or absence of disease, and a set of covariates related to the outcome through linear effects, interactions, and nonlinear relationships. Because the data-generating mechanism is known, the behavior of SL can be evaluated under a controlled and predefined structure. \\
A total of $n=500$ observations and five covariates $(\mathbf{W}_1,\ldots,\mathbf{W}_5)$ are generated. The first four covariates are drawn from uniform distributions and illustratively represent standardized characteristics such as age, body mass index, blood pressure, and a biomarker. The fifth covariate, $\mathbf{W}_5$, represents a binary comorbidity indicator whose probability depends on $\mathbf{W}_2$ and $\mathbf{W}_3$.

\begin{rcode}[label={code:sim-covariates}]{Generating simulated covariates}
	set.seed(2743, kind = "Mersenne-Twister")
	n <- 500
	
	# Covariates (standardised to unit scale)
	W1 <- runif(n, 0.5, 1)
	W2 <- runif(n, 0, 1)
	W3 <- runif(n, 0.25, 0.75)
	W4 <- runif(n, 0, 1)
	
	# A derived comorbidity indicator
	W5 <- rbinom(n, 1, 1 / (1 + exp(1.5 * W2 - W3)))
\end{rcode}

The binary outcome $Y$ is generated using a logistic model that includes main effects, an interaction between $\mathbf{W}_5$ and $\mathbf{W}_1$, and a nonlinear term involving $\mathbf{W}_4$. Consequently, the data-generating mechanism represents a structure that is more complex than a purely linear relationship between the covariates and the outcome.

\begin{rcode}[label={code:sim-outcome}]{Generating the binary outcome}
	# True data-generating mechanism: interactions + non-linearity
	logit_p <- -0.2 * W5 - 2 * W1 + 4 * W5 * W1 - 1.5 * W2 + sin(W4 * pi)
	Y <- rbinom(n, 1, 1 / (1 + exp(-logit_p)))
	
	dat <- data.frame(W1, W2, W3, W4, W5, Y)
\end{rcode}

Once the data have been generated, the resulting distribution of the simulated sample can be examined. Because the outcome $Y$ is binary, its mean corresponds to the proportion of observations for which $Y=1$, and therefore provides the prevalence of the simulated outcome. In addition, \texttt{summary()} provides an initial descriptive summary of the generated variables.

\begin{rcode}[label={code:descriptive}]{Descriptive summary of the simulated data}
	cat("Prevalence of the outcome:", round(mean(Y), 3), "\n")
	summary(dat)
\end{rcode}

\routputcaption{box:summary}{Simulated data descriptive summary.}
\begin{routputboxref}[label={box:summary}]
	\centering
	\ttfamily
	\resizebox{\linewidth}{!}{%
		\begin{tabular}{lrrrrrr}
			& W1     & W2      & W3     & W4      & W5    & Y    \\
			Min.      & 0.500  & 0.00056 & 0.252  & 0.00388 & 0.000 & 0.00 \\
			1st Qu.   & 0.615  & 0.24025 & 0.372  & 0.24555 & 0.000 & 0.00 \\
			Median    & 0.728  & 0.47032 & 0.502  & 0.49973 & 0.000 & 0.00 \\
			Mean      & 0.740  & 0.47600 & 0.495  & 0.50607 & 0.436 & 0.44 \\
			3rd Qu.   & 0.864  & 0.72776 & 0.618  & 0.76426 & 1.000 & 1.00 \\
			Max.      & 0.999  & 0.99765 & 0.750  & 0.99940 & 1.000 & 1.00 \\
		\end{tabular}%
	}
\end{routputboxref}

To further examine the distribution of the simulated binary outcome, a bar plot can be used as an additional descriptive tool:
\begin{figure}
	\centering
	\includegraphics[width=\linewidth]{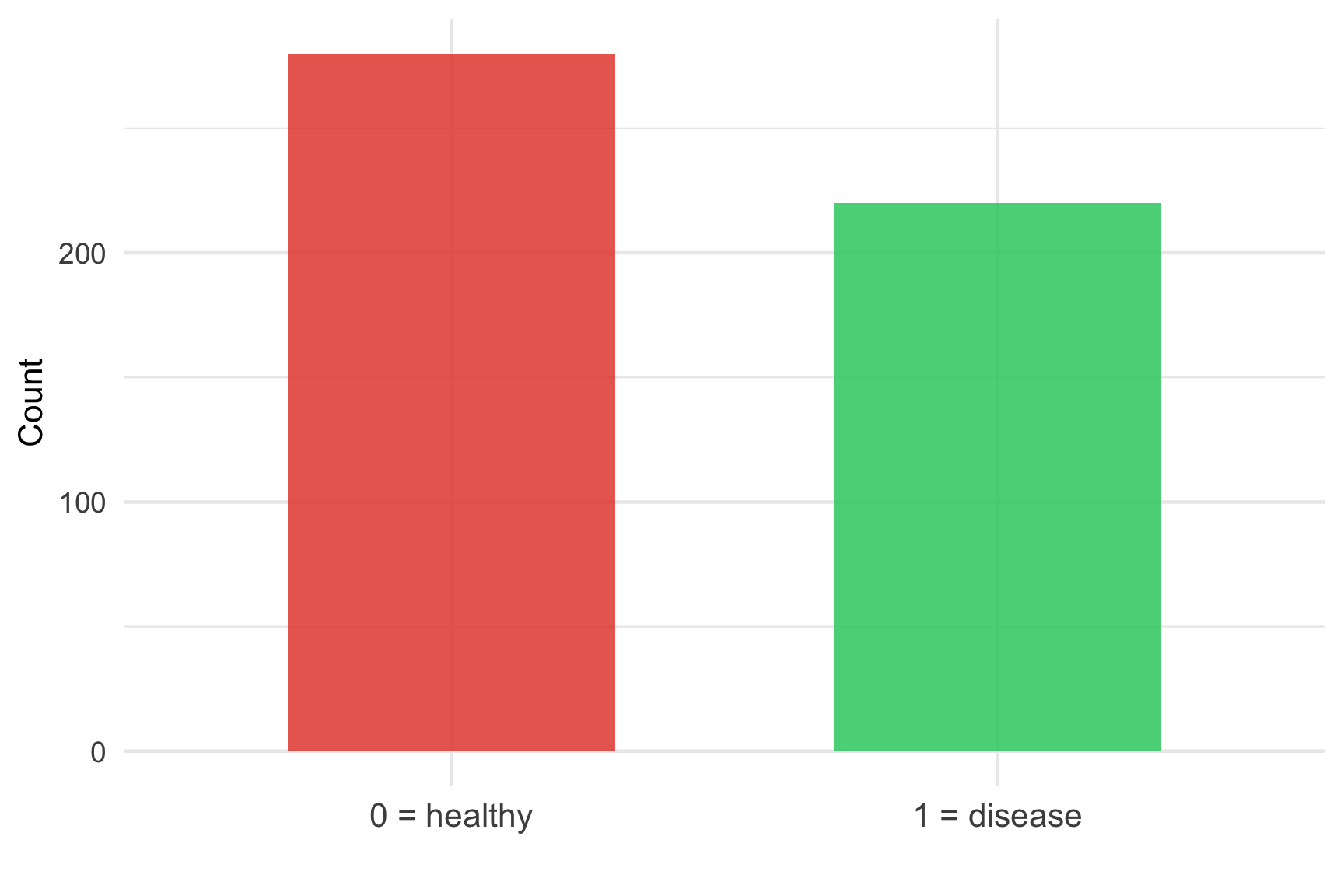}
	\caption{Distribution of the simulated binary outcome.}
	\label{fig:outcome_distribution}
\end{figure}
A total of 44\% of the simulated observations have the outcome of interest,  corresponding to the presence of disease ($Y=1$), while the covariates remain within the ranges specified during data generation. These data will subsequently be used to progressively evaluate the performance of different candidate  algorithms and their combination through SL.

\section{Basic SL}

The first practical approach to SL begins with a single candidate algorithm to familiarize the reader with the \texttt{SuperLearner()} function and the main elements of its output. Additional algorithms are subsequently incorporated  into the library to examine how SL assigns weights among the candidates according to their cross-validated performance.

\subsection{A Single Algorithm}

The first fit uses penalized logistic regression through LASSO/elastic net, implemented in SL using \texttt{SL.glmnet}.\cite{Tibshirani1996,Friedman2010} The simulated data are divided into two sets: 80\% of the observations are used for training, and the remaining 20\% are reserved for testing. To preserve the same partition in subsequent analyses, the random seed is fixed using \texttt{set.seed(123)} (Box~\ref{code:sl-single}).

\begin{rcode}[label={code:sl-single}]{Fitting SL with a single algorithm}
	set.seed(
		123,
		kind = "Mersenne-Twister",
		normal.kind = "Inversion",
		sample.kind = "Rejection"
	)
	
	# Train / test split (80 % / 20 %)
	train_idx <- sample(1:n, 400)
	X_train   <- dat[train_idx, 1:5]
	Y_train   <- dat[train_idx, 6]
	X_test    <- dat[-train_idx, 1:5]
	Y_test    <- dat[-train_idx, 6]
	
	sl_lasso <- SuperLearner(
	Y = Y_train, X = X_train,
	family = binomial(),
	SL.library = "SL.glmnet"
	)
	sl_lasso
\end{rcode}

The output of \texttt{sl\_lasso} reports two main elements: the risk estimated through  CV (\texttt{Risk}) and the weight assigned to the algorithm (\texttt{Coef}).  In this first fit, the weight is equal to 1 because the library contains only  \texttt{SL.glmnet}, and therefore there are no other algorithms among which the weight can be distributed. Although these results are presented together in the general output, they can also be accessed separately using \texttt{sl\_lasso\$cvRisk} and \texttt{sl\_lasso\$coef}.

\routputcaption{box:sl-single}{SL fit with a single candidate algorithm.}
\begin{routputboxref}[label={box:sl-single}]
	\ttfamily
	\raggedright
	Call:
	SuperLearner(
	Y = Y\_train,
	X = X\_train,
	family = binomial(),
	SL.library = "SL.glmnet"
	)
	\medskip
	\centering
	\begin{tabular}{lrr}
		& Risk & Coef \\
		SL.glmnet\_All & 0.1612 & 1 \\
	\end{tabular}
\end{routputboxref}

To examine the model fitted by \texttt{SL.glmnet}, the coefficients corresponding to each covariate can be extracted at the penalty value selected by CV  (\texttt{lambda.min}) using Box~\ref{code:lasso-coef}:

\begin{rcode}[label={code:lasso-coef}]{Extracting LASSO coefficients}
	# The fitted glmnet object is stored in the fitLibrary
	lasso_fit <- sl_lasso$fitLibrary$SL.glmnet_All$object
	round(as.matrix(coef(lasso_fit, s = "lambda.min")), 3)
\end{rcode}

The resulting output is:

\routputcaption{box:lasso-coefficients}{Coefficients estimated by \texttt{SL.glmnet} at \texttt{lambda.min}.}
\begin{routputboxref}[label={box:lasso-coefficients}]
	\ttfamily
	\centering
	\begin{tabular}{lr}
		& lambda.min \\
		(Intercept) & -1.478 \\
		W1          &  0.183 \\
		W2          & -0.778 \\
		W3          &  0.004 \\
		W4          &  0.652 \\
		W5          &  2.545 \\
	\end{tabular}
\end{routputboxref}

The result of this first approach will serve as a reference for comparing the changes observed as additional algorithms are incorporated into the library.

\subsection{Two Algorithms}
For the second fit (Box~\ref{code:sl-two}), \texttt{SL.randomForest} is added to the library while retaining the algorithm used in the previous case (\texttt{SL.glmnet}).\cite{Breiman2001,Liaw2002} SL therefore estimates a combination of both candidates and assigns a weight to each according to its performance within the CV procedure.

\begin{rcode}[label={code:sl-two}]{Fitting SL with two algorithms}
	set.seed(
		123,
		kind = "Mersenne-Twister",
		normal.kind = "Inversion",
		sample.kind = "Rejection"
	)
	
	sl_two <- SuperLearner(
	Y = Y_train, X = X_train,
	family = binomial(),
	SL.library = c("SL.glmnet", "SL.randomForest")
	)
	sl_two
\end{rcode}

\routputcaption{box:sl-two}{SL fit with two candidate algorithms.}
\begin{routputboxref}[label={box:sl-two}]
	\ttfamily
	\raggedright
	Call:
	SuperLearner(
	Y = Y\_train, X = X\_train, family = binomial(),
	SL.library = c("SL.glmnet", "SL.randomForest")
	)
	\medskip
	\centering
	\begin{tabular}{lrr}
		& Risk & Coef \\
		SL.glmnet\_All       & 0.1593 & 0.6851 \\
		SL.randomForest\_All & 0.1655 & 0.3149 \\
	\end{tabular}
\end{routputboxref}

After incorporating a second algorithm, the ensemble assigns nonzero weights to both candidates. Whereas \texttt{SL.glmnet} received a weight of 1 in the single-algorithm fit, its weight decreases to $0.6851$ in the two-algorithm ensemble, while \texttt{SL.randomForest} receives a weight of $0.3149$. The cross-validated risk is lower for \texttt{SL.glmnet} ($0.1593$) than for \texttt{SL.randomForest} ($0.1655$). Nevertheless, both algorithms contribute to the final weighted prediction, with a larger contribution assigned to \texttt{SL.glmnet}.

\subsection{A Full Ensemble with a Reference Model}
A reference model is a simple model used as a point of comparison to assess whether more complex models provide an improvement in predictive performance. In this case, \texttt{SL.mean} is added to the library. This learner predicts the sample mean of the outcome. Because the outcome is binary, this prediction corresponds to the observed prevalence and is the same for all individuals. Owing to its simplicity and low computational cost, it provides a useful reference for assessing the contribution of the other algorithms in the library.

\begin{rcode}[label={code:sl-reference}]{Including a reference model in the ensemble}
	set.seed(
		123,
		kind = "Mersenne-Twister",
		normal.kind = "Inversion",
		sample.kind = "Rejection"
	)
	sl <- SuperLearner(
	Y = Y_train, X = X_train,
	family = binomial(),
	SL.library = c("SL.mean", "SL.glmnet", "SL.randomForest")
	)
	sl
\end{rcode}

\routputcaption{box:sl-reference}{SL fit including the reference model.}
\begin{routputboxref}[label={box:sl-reference}]
	\ttfamily
	\raggedright
	Call:
	SuperLearner(
	Y = Y\_train, X = X\_train, family = binomial(),
	SL.library = c("SL.mean", "SL.glmnet", "SL.randomForest")
	)
	\medskip
	\centering
	\begin{tabular}{lrr}
		& Risk & Coef \\
		SL.mean\_All         & 0.2478 & 0.0000 \\
		SL.glmnet\_All       & 0.1593 & 0.6851 \\
		SL.randomForest\_All & 0.1655 & 0.3149 \\
	\end{tabular}
\end{routputboxref}

The inclusion of \texttt{SL.mean} provides a simple reference against which the other candidate algorithms can be compared. Because this learner predicts the observed outcome prevalence for every individual, it does not account for differences in covariate profiles. Its cross-validated risk ($0.2478$) is higher than that of \texttt{SL.glmnet} ($0.1593$) and \texttt{SL.randomForest} ($0.1655$). In the resulting ensemble, the meta-learner assigns a weight of $0$ to \texttt{SL.mean}, while retaining weights of $0.6851$ and $0.3149$ for \texttt{SL.glmnet} and \texttt{SL.randomForest}, respectively. These weights are identical to those obtained with the two-algorithm library, illustrating how SL can incorporate a reference model while selecting the combination of predictions that minimizes the cross-validated loss.

\begin{rcode}[label={code:sl-time}]{Checking computational time}
	# Time it took to fit
	sl$times$everything
\end{rcode}

\routputcaption{box:sl-time}{Computational time required to fit the SL ensemble.}
\begin{routputref}[label={box:sl-time}]
	user  system elapsed
	4.993   0.184   5.219
\end{routputref}

\section{Predictions and Performance}

Once the ensemble has been fitted and the weights of its component algorithms have been estimated, the 20\% of the data previously reserved as a test set is used to generate predictions and evaluate predictive performance.

\subsection{Generating Predictions}

The \texttt{predict()} function uses the fitted ensemble to generate predictions for new data. In this case, both the final SL prediction and the individual predictions generated by each algorithm in the library are retained. To obtain both types of predictions, the argument \texttt{onlySL = FALSE} is specified.

\begin{rcode}[label={code:predict}]{Generating predictions on the test set}
	# Keep ensemble and individual learner predictions
	pred <- predict(sl, newdata = X_test, onlySL = FALSE)
	str(pred)
\end{rcode}

\routputcaption{box:prediction-structure}{Structure of the prediction object returned by \texttt{predict()}.}
\begin{routputref}[label={box:prediction-structure}]
	List of 2
	$ pred           : num [1:100, 1] 0.191 0.156 0.182 0.704 0.248 ...
	$ library.predict: num [1:100, 1:3] 0.44 0.44 0.44 0.44 0.44 0.44 0.44 0.44 0.44 0.44 ...
	..- attr(*, "dimnames")=List of 2
	.. ..$ : NULL
	.. ..$ : chr [1:3] "SL.mean_All" "SL.glmnet_All" "SL.randomForest_All"
\end{routputref}

The resulting object contains two components. \texttt{pred\$pred} stores the final ensemble predictions, whereas \texttt{pred\$library.predict} contains the predictions generated by each candidate algorithm. In this case, 100 predictions are generated, corresponding to the observations included in the test set.

The first ensemble predictions can be examined using (Box~\ref{code:head-pred}):

\begin{rcode}[label={code:head-pred}]{Examining ensemble predictions}
	# Ensemble predictions (probability of disease)
	head(pred$pred)
\end{rcode}

\routputcaption{box:ensemble-predictions}{First predicted probabilities generated by the SL ensemble.}
\begin{routputboxref}[label={box:ensemble-predictions}]
	\ttfamily
	\centering
	\begin{tabular}{lr}
		& {[,1]} \\
		{[1,]} & 0.1914 \\
		{[2,]} & 0.1561 \\
		{[3,]} & 0.1824 \\
		{[4,]} & 0.7044 \\
		{[5,]} & 0.2478 \\
		{[6,]} & 0.1648 \\
	\end{tabular}
\end{routputboxref}

Each value represents the probability of the outcome predicted by the ensemble for an individual in the test set. For example, the first individual receives a predicted probability of approximately $0.191$, or $19.1\%$. These probabilities are obtained by combining the predictions of the candidate algorithms according to their estimated ensemble weights. Consequently, individuals with different covariate profiles may receive different predicted probabilities.

\begin{rcode}[label={code:head-library}]{Examining individual learner predictions}
	# Predictions from each individual learner
	head(pred$library.predict)
\end{rcode}

\routputcaption{box:individual-predictions}{First predicted probabilities generated by the individual candidate algorithms.}
\begin{routputboxref}[label={box:individual-predictions}]
	\ttfamily
	\centering
	\resizebox{\linewidth}{!}{%
		\begin{tabular}{lrrr}
			& SL.mean\_All & SL.glmnet\_All & SL.randomForest\_All \\
			{[1,]} & 0.44 & 0.2532 & 0.057 \\
			{[2,]} & 0.44 & 0.1690 & 0.128 \\
			{[3,]} & 0.44 & 0.1913 & 0.163 \\
			{[4,]} & 0.44 & 0.7569 & 0.590 \\
			{[5,]} & 0.44 & 0.2142 & 0.321 \\
			{[6,]} & 0.44 & 0.2198 & 0.045 \\
		\end{tabular}%
	}
\end{routputboxref}

\subsection{Manual Weighted Sum: Understanding the Ensemble Prediction}
\label{sec:manual-weighted-sum}

The \texttt{predict()} function automatically computes the weighted sum of individual learner predictions. To illustrate the mechanics, Box~\ref{code:manual-prediction} shows how to manually reconstruct the ensemble prediction using the fitted SL weights and the individual learner predictions.

\begin{rcode}[label={code:manual-prediction}]{Manual weighted sum reconstruction}
	# Extract fitted ensemble weights from the SL object
	alpha_sl <- sl$coef
	names(alpha_sl) <- c("SL.mean", "SL.glmnet", "SL.randomForest")
	print(round(alpha_sl, 4))
	
	# Extract individual learner predictions for test set
	lib_pred <- pred$library.predict
	colnames(lib_pred) <- c("SL.mean", "SL.glmnet", "SL.randomForest")
	
	# Manual weighted sum: psi_SL = sum(alpha_k * psi_k)
	manual_sl_pred <- as.vector(lib_pred %*% alpha_sl)
	
	# Compare with predict() output
	ensemble_pred <- pred$pred
	
	# Show first 6 predictions
	data.frame(
	SL.mean = round(lib_pred[1:6, "SL.mean"], 3),
	SL.glmnet = round(lib_pred[1:6, "SL.glmnet"], 3),
	SL.randomForest = round(lib_pred[1:6, "SL.randomForest"], 3),
	Manual_SL = round(manual_sl_pred[1:6], 3),
	Auto_SL = round(ensemble_pred[1:6], 3),
	Match = round(manual_sl_pred[1:6], 6) == round(ensemble_pred[1:6], 6)
	)
\end{rcode}

\routputcaption{box:manual-prediction}{Comparison of manually reconstructed and automatically generated SL predictions.}
\begin{routputboxref}[label={box:manual-prediction}]
	\ttfamily
	\centering
	\resizebox{\linewidth}{!}{%
		\begin{tabular}{lrrrrrl}
			& SL.mean & SL.glmnet & SL.randomForest & Manual\_SL & Auto\_SL & Match \\
			1 & 0.440 & 0.253 & 0.057 & 0.191 & 0.191 & TRUE \\
			2 & 0.440 & 0.169 & 0.128 & 0.156 & 0.156 & TRUE \\
			3 & 0.440 & 0.191 & 0.163 & 0.182 & 0.182 & TRUE \\
			4 & 0.440 & 0.757 & 0.590 & 0.704 & 0.704 & TRUE \\
			5 & 0.440 & 0.214 & 0.321 & 0.248 & 0.248 & TRUE \\
			6 & 0.440 & 0.220 & 0.045 & 0.165 & 0.165 & TRUE \\
		\end{tabular}%
	}
\end{routputboxref}

\begin{figure}
	\centering
	\includegraphics[width=\linewidth]{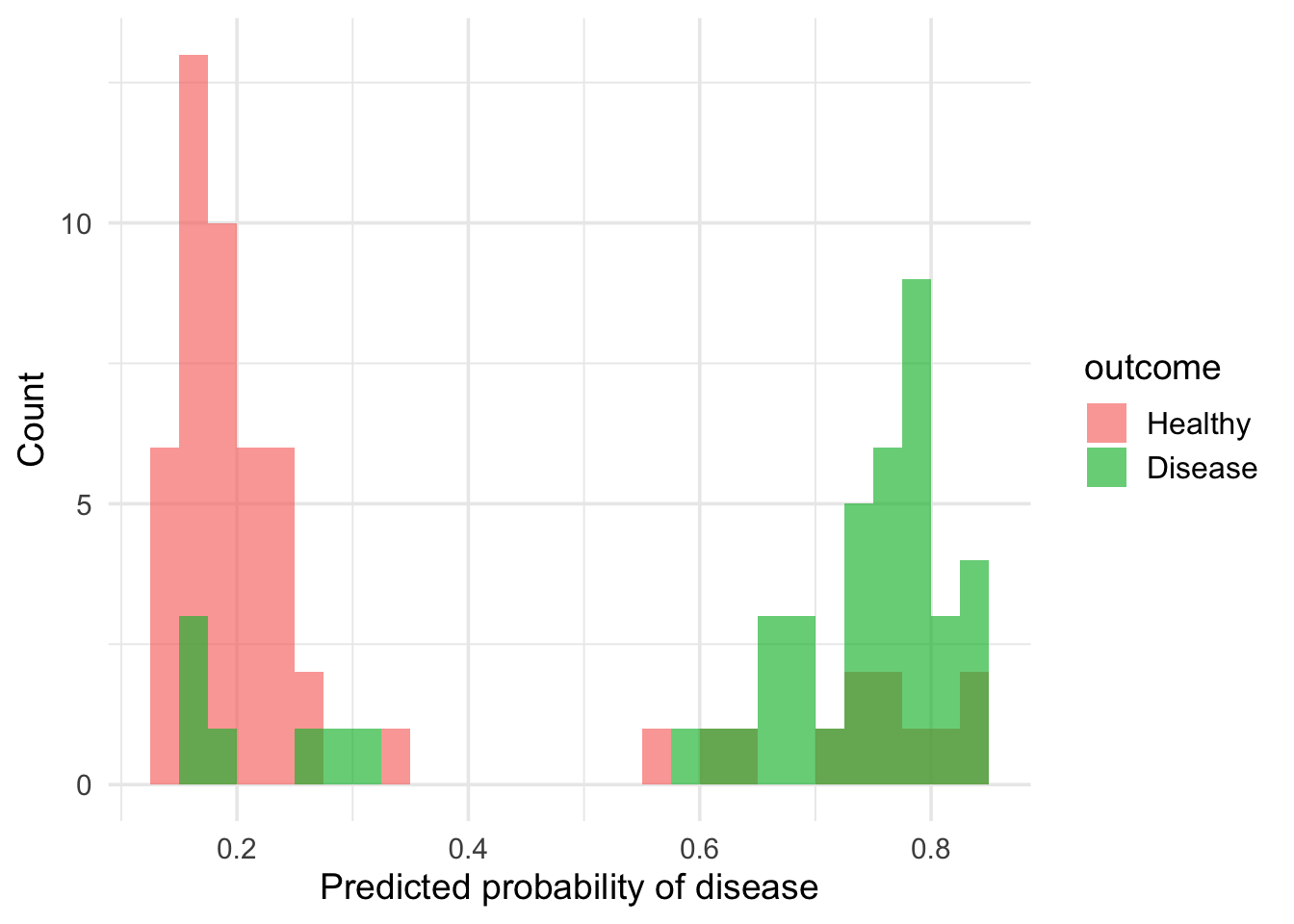}
	\caption{Predicted probabilities by true outcome.}
	\label{fig:predicted_probabilities}
\end{figure}

The ensemble prediction is obtained by multiplying each candidate algorithm's predicted probability by its estimated weight and summing the resulting values. In this example, the estimated weights are $\alpha = (0.0000, 0.6851, 0.3149)$ for \texttt{SL.mean}, \texttt{SL.glmnet}, and \texttt{SL.randomForest}, respectively. Because \texttt{SL.mean} receives a weight of zero, the final prediction is determined by the weighted combination of LASSO and Random Forest. The manually reconstructed probabilities agree with those returned by \texttt{predict()} up to numerical precision, as confirmed by the \texttt{Match} column. This equivalence illustrates how the estimated ensemble weights are used to generate predictions for new observations. \\
To further examine the predictions, Figure~\ref{fig:predicted_probabilities} displays the distribution of the estimated probabilities according to the observed disease status in the test set.

The predicted probabilities are generally lower for individuals without disease and higher for those with disease, indicating that the ensemble assigns different levels of predicted risk according to the observed outcome groups. Although the distributions show a clear separation, some overlap remains, illustrating that individuals with and without disease cannot always be distinguished by a single probability threshold. This graphical representation provides an intuitive assessment of the ensemble's discriminatory ability, which is subsequently quantified using the ROC curve and the area under the curve (AUC).

\subsection{Area Under the ROC Curve}

The discriminatory ability of SL can be summarized using the area under the ROC curve (AUC). This measure evaluates the ability of the model to assign higher predicted probabilities to individuals with the outcome than to those without the outcome. Values close to $0.5$ indicate limited discrimination, whereas values approaching $1$ reflect greater discriminatory ability.\cite{HanleyMcNeil1982,Steyerberg2010}

In this case, the AUC is calculated using the predictions obtained on the test set (Box~\ref{code:auc}):\cite{Sing2005ROCR}

\begin{rcode}[label={code:auc}]{Computing the AUC on the test set}
	pred_rocr <- ROCR::prediction(pred$pred, Y_test)
	
	auc_sl <- ROCR::performance(
	pred_rocr,
	measure = "auc",
	x.measure = "cutoff"
	)@y.values[[1]]
	
	cat("Test AUC:", round(auc_sl, 4), "\n")
\end{rcode}

\routputcaption{box:test-auc}{Area under the ROC curve for the SL ensemble on the test set.}
\begin{routputref}[label={box:test-auc}]
	Test AUC: 0.8413
\end{routputref}

SL achieves an AUC of $0.8413$ on the held-out test set. This value indicates that the ensemble has good discriminatory ability, tending to assign higher predicted probabilities to individuals with the outcome than to those without it. \\
To compare the ensemble with the individual algorithms, the AUC and ROC curve are also calculated for each candidate in the library (Box~\ref{code:roc-curves}).\cite{Robin2011pROC} This allows the performance of \texttt{SL.mean}, \texttt{SL.glmnet}, \texttt{SL.randomForest}, and SL to be examined jointly.

\begin{rcode}[label={code:roc-curves}]{ROC curves for the ensemble and individual algorithms}
	# Column order of library.predict matches SL.library:
	#  1 = SL.mean, 2 = SL.glmnet, 3 = SL.randomForest
	auc_mean   <- pROC::auc(Y_test, pred$library.predict[, 1])
	auc_glmnet <- pROC::auc(Y_test, pred$library.predict[, 2])
	auc_rf     <- pROC::auc(Y_test, pred$library.predict[, 3])
	
	roc_mean   <- pROC::roc(Y_test, pred$library.predict[, 1], quiet = TRUE)
	roc_glmnet <- pROC::roc(Y_test, pred$library.predict[, 2], quiet = TRUE)
	roc_rf     <- pROC::roc(Y_test, pred$library.predict[, 3], quiet = TRUE)
	roc_sl     <- pROC::roc(Y_test, as.numeric(pred$pred), quiet = TRUE)
	
	roc_df <- data.frame(
	FPR   = c(1 - roc_mean$specificities, 1 - roc_glmnet$specificities,
	1 - roc_rf$specificities, 1 - roc_sl$specificities),
	TPR   = c(roc_mean$sensitivities, roc_glmnet$sensitivities,
	roc_rf$sensitivities, roc_sl$sensitivities),
	Model = rep(c(
	paste0("Mean (AUC = ", round(auc_mean, 3), ")"),
	paste0("LASSO (AUC = ", round(auc_glmnet, 3), ")"),
	paste0("Random Forest (AUC = ", round(auc_rf, 3), ")"),
	paste0("SuperLearner (AUC = ", round(auc_sl, 3), ")")
	), times = sapply(list(roc_mean, roc_glmnet, roc_rf, roc_sl),
	function(x) length(x$sensitivities)))
	)
	
	ggplot(roc_df, aes(x = FPR, y = TPR, color = Model)) +
	geom_path(linewidth = 1.1) +
	geom_abline(intercept = 0, slope = 1, linetype = "dashed",
	color = "gray50") +
	scale_color_manual(values = c("#3498DB", "#E74C3C", "#9B59B6", "#2ECC71")) +
	labs(x = "1 - Specificity", y = "Sensitivity") +
	xlim(0, 1) + ylim(0, 1) +
	theme_minimal(base_size = 14) +
	theme(legend.position = "inside", legend.position.inside = c(0.65, 0.25))
\end{rcode}

\begin{figure}
	\centering
	\includegraphics[width=\linewidth]{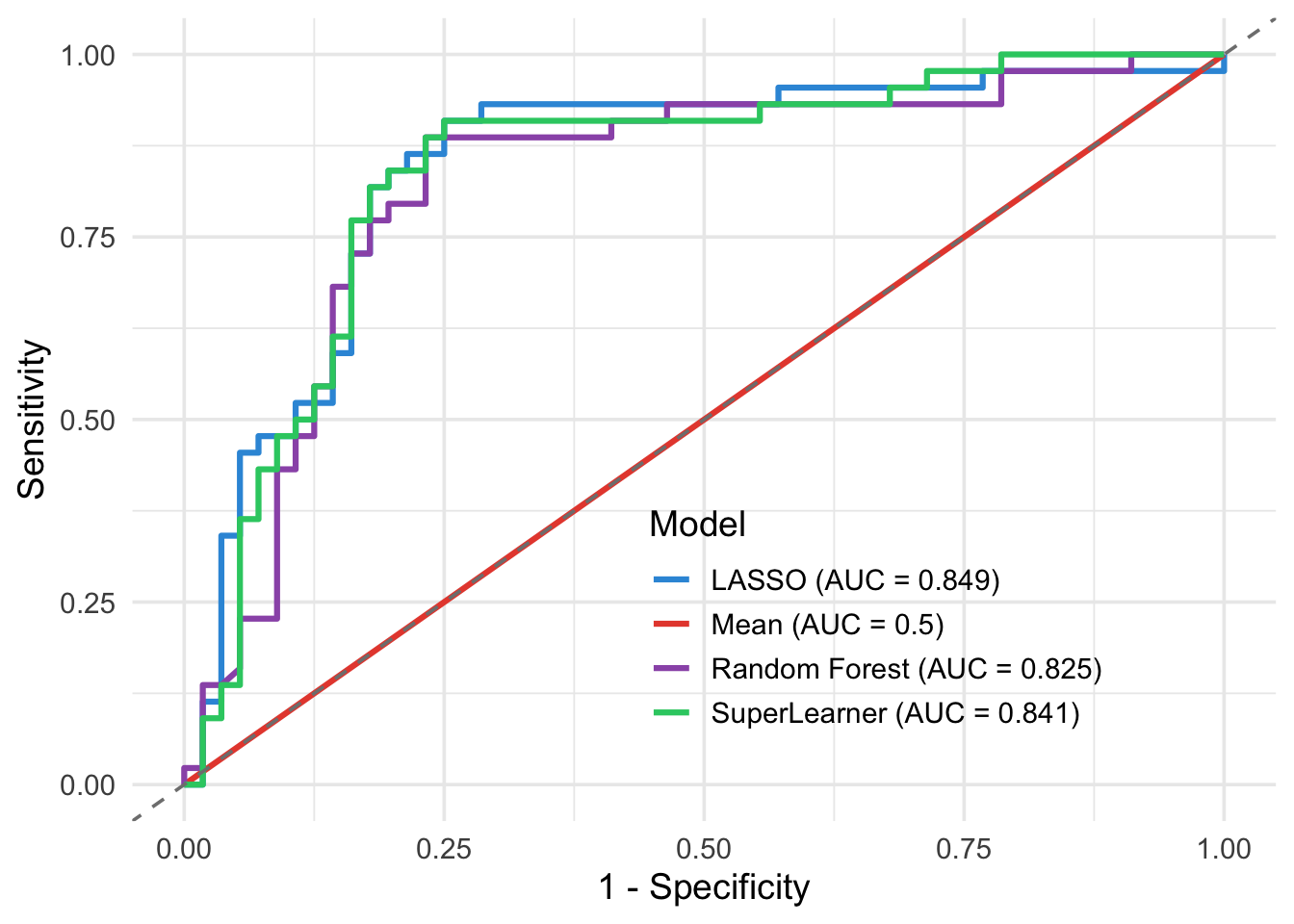}
	\caption{ROC curves for SL and the individual candidate algorithms.}
	\label{fig:pla_mul}
\end{figure}

Figure~\ref{fig:pla_mul} shows the differences in discriminatory performance among the evaluated models. Because \texttt{SL.mean} generates a constant prediction for all observations, it lies on the reference line. In contrast, \texttt{SL.glmnet}, \texttt{SL.randomForest}, and SL show greater ability to distinguish between individuals with and without the outcome. \\
In this test set, \texttt{SL.glmnet} achieves the highest observed AUC ($0.849$), followed by the SL ensemble ($0.841$) and \texttt{SL.randomForest} ($0.825$). Although SL combines predictions from multiple candidate algorithms, it does not necessarily attain the highest AUC in every test sample. In this example, the ensemble weights were estimated by minimizing cross-validated squared-error loss rather than directly maximizing the AUC. The results therefore illustrate the distinction between optimizing a specified loss function during training and evaluating discrimination on independent test data.

\section{Cross-Validated SL}
Up to this point, the performance of SL has been evaluated using a single training-test split. Although this strategy allows the model to be assessed on observations that were not used during fitting, the resulting estimates correspond to one specific partition of the data. CV complements this analysis by evaluating the procedure across different partitions and summarizing both its average performance and its variability.\cite{VarmaSimon2006} \\
In this context, a distinction should be made between the internal CV used to estimate the risk of the candidate algorithms and determine the ensemble weights, and the external CV implemented through \texttt{CV.SuperLearner()}, which is used to evaluate the complete procedure.\cite{CawleyTalbot2010} In each external partition, SL is refitted using the corresponding training data and evaluated on observations that were not used in that fit.

\subsection{Fitting \texttt{CV.SuperLearner}} 
In this example, the 400 observations in the training set are divided into five partitions of 80 observations each. At each iteration, 320 observations are used to refit SL, including its internal CV procedure, while the remaining 80 observations are used to evaluate its risk. The process is repeated until each partition has served once as the external evaluation set.

\begin{rcode}[label={code:cvsl}]{Fitting CV.SuperLearner}
	set.seed(123)
	
	system.time({
		cv_sl <- CV.SuperLearner(
		Y = Y_train, X = X_train,
		family = binomial(),
		V = 5,
		SL.library = c("SL.mean", "SL.glmnet", "SL.randomForest")
		)
	})
\end{rcode}

\routputcaption{box:cvsl-time}{Computational time required to fit \texttt{CV.SuperLearner()}.}
\begin{routputref}[label={box:cvsl-time}]
	user  system elapsed
	18.363   1.025  20.128
\end{routputref}

The summary of the \texttt{cv\_sl} object allows the risk obtained across the five external partitions to be examined for both SL and the individual algorithms in the library:

\begin{rcode}[label={code:cvsl-summary}]{Summarizing CV.SuperLearner results}
	summary(cv_sl)
\end{rcode}

\routputcaption{box:cvsl-summary}{Cross-validated risk summary for SL and the candidate algorithms.}
\begin{routputboxref}[label={box:cvsl-summary}]
	\ttfamily
	Call:\par
	CV.SuperLearner(Y = Y\_train, X = X\_train, V = 5, family = binomial(),
	SL.library = c("SL.mean", "SL.glmnet", "SL.randomForest"))\par
	\medskip
	Risk is based on: Mean Squared Error\par
	\medskip
	All risk estimates are based on V = 5\par
	\medskip
	\centering
	\begin{tabular}{lrrrr}
		Algorithm & Ave & se & Min & Max \\
		Super Learner       & 0.16 & 0.0115 & 0.13 & 0.21 \\
		Discrete SL         & 0.16 & 0.0116 & 0.13 & 0.22 \\
		SL.mean\_All        & 0.25 & 0.0031 & 0.24 & 0.27 \\
		SL.glmnet\_All      & 0.16 & 0.0116 & 0.13 & 0.22 \\
		SL.randomForest\_All & 0.17 & 0.0119 & 0.15 & 0.22 \\
	\end{tabular}
\end{routputboxref}

\begin{figure}
	\centering
	\includegraphics[width=\linewidth]{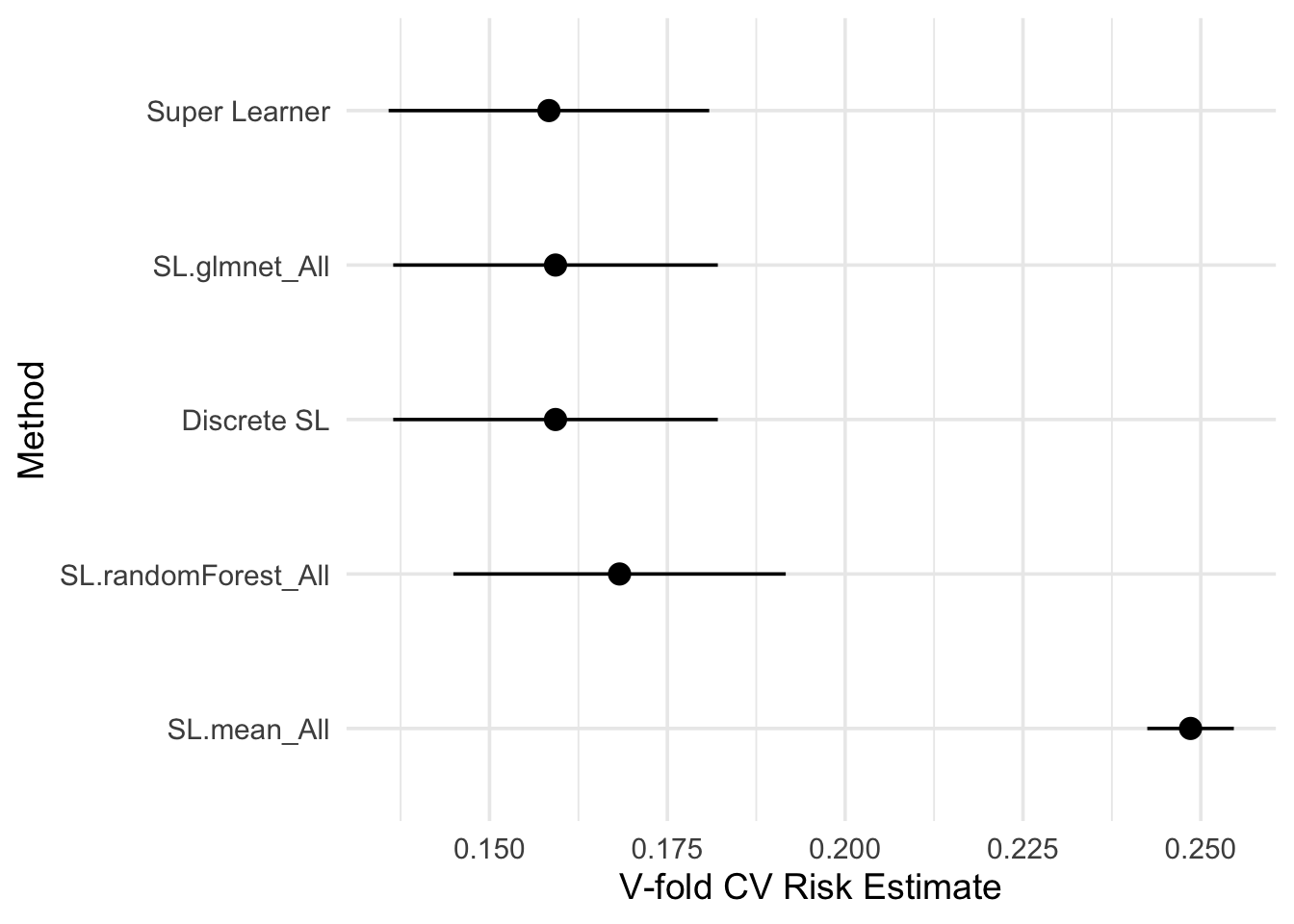}
	\caption{Cross-validated risk (mean squared error) with 95\% confidence intervals.}
	\label{fig:cv_risk}
\end{figure}

In this case, risk is expressed using the mean squared error. The \texttt{Ave} column corresponds to the average risk obtained across the five external partitions, \texttt{se} to its standard error, and \texttt{Min} and \texttt{Max} to the observed range across partitions. The \textit{Super Learner} row summarizes the performance of the full ensemble, whereas \textit{Discrete SL} corresponds to the best individual candidate algorithm; this comparison makes it possible to assess whether the weighted combination provides an improvement over selecting a single candidate. \\
The SL ensemble achieves an average cross-validated risk of $0.15835$, slightly lower than that of Discrete SL and \texttt{SL.glmnet} ($0.15928$). \texttt{SL.randomForest} has a higher average risk ($0.16828$), while \texttt{SL.mean} has the highest value ($0.24856$). Because risk is measured using mean squared error, lower values indicate better predictive performance under this criterion. The small difference between SL and Discrete SL illustrates how combining predictions can produce a different cross-validated risk from selecting a single candidate algorithm. Figure~\ref{fig:cv_risk} displays these estimates together with their 95\% confidence intervals.

Figure~\ref{fig:cv_risk} shows that the risk estimates for SL, Discrete SL, and \texttt{SL.glmnet} are concentrated at very similar values, whereas \texttt{SL.mean} is clearly shifted toward higher risk. The intervals for the methods with lower risk show substantial overlap, although this pattern should not by itself be interpreted as formal evidence of equivalent performance. \\
To identify which candidate algorithm corresponds to the Discrete SL in each outer fold, the object \texttt{cv\_sl\$whichDiscreteSL} can be examined (Box~\ref{code:discrete-sl}):

\begin{rcode}[label={code:discrete-sl}]{Identifying the Discrete SL}
	table(simplify2array(cv_sl$whichDiscreteSL))
\end{rcode}

\routputcaption{box:discrete-sl-selection}{Best individual candidate selected across the external CV folds.}
\begin{routputref}[label={box:discrete-sl-selection}]
	SL.glmnet_All
	5
\end{routputref}

In all five external folds, \texttt{SL.glmnet} was selected as the Discrete SL based on the candidate algorithms' internal cross-validation results. Consequently, Discrete SL and \texttt{SL.glmnet} have identical external cross-validated risk estimates. This example illustrates the distinction between Discrete SL, which selects a single candidate algorithm within each training fold, and the SL ensemble, which combines predictions using estimated weights. The contribution of each candidate to the weighted ensemble can be examined through the distribution of weights across the external folds.

\subsection{Distribution of Ensemble Weights}
In addition to identifying the best individual candidate, it is useful to examine how the ensemble weights vary across the external CV folds. This makes it possible to assess whether the weighting pattern observed in a single fit remains similar when the composition of the training data changes.

\begin{rcode}[label={code:review-weights}]{Distribution of ensemble weights across folds}
	review_weights <- function(cv_sl) {
		meta_weights <- coef(cv_sl)              # matrix: folds x learners
		means <- colMeans(meta_weights)
		sds   <- apply(meta_weights, 2, sd)
		mins  <- apply(meta_weights, 2, min)
		maxs  <- apply(meta_weights, 2, max)
		stats <- cbind("mean(weight)" = means, sd = sds, min = mins, max = maxs)
		stats[order(stats[, 1], decreasing = TRUE), ]
	}
	print(review_weights(cv_sl), digits = 3)
\end{rcode}

\routputcaption{box:weight-distribution}{Distribution of ensemble weights across the external CV folds.}
\begin{routputboxref}[label={box:weight-distribution}]
	\ttfamily
	\centering
	\resizebox{\linewidth}{!}{%
		\begin{tabular}{lrrrr}
			Algorithm & mean(weight) & sd & min & max \\
			SL.glmnet\_All       & 0.768 & 0.0499 & 0.694 & 0.819 \\
			SL.randomForest\_All & 0.232 & 0.0499 & 0.181 & 0.306 \\
			SL.mean\_All         & 0.000 & 0.0000 & 0.000 & 0.000 \\
		\end{tabular}%
	}
\end{routputboxref}

As shown in Box~\ref{box:weight-distribution}, \texttt{SL.glmnet} receives the largest average weight ($0.768$), followed by \texttt{SL.randomForest} ($0.232$), whereas \texttt{SL.mean} receives a weight of $0$ in all folds. This pattern is consistent with the previous results, in which \texttt{SL.glmnet} showed the best performance among the individual candidate algorithms. \\
The variation in the weights across folds is relatively small. For \texttt{SL.glmnet}, the weights range from $0.694$ to $0.819$, whereas for \texttt{SL.randomForest} they range from $0.181$ to $0.306$. In this example, the distribution of the weights remains broadly similar across the five external CV folds.

\section{Hyperparameter Tuning}

Up to this point, each algorithm has been incorporated into the library using a single configuration. However, many algorithms depend on hyperparameters that can modify their predictive behavior. In SL, different configurations of the same algorithm can be incorporated as distinct candidates and evaluated using the same CV procedure introduced previously.\cite{Phillips2023}

\subsection{Algorithm Variants}

Random Forest provides a useful example of this possibility.\cite{Breiman2001,Probst2019} One of its hyperparameters is \texttt{ntree}, which determines the number of trees used. Until now, \texttt{SL.randomForest} has been used with its default configuration. To evaluate an alternative with 3000 trees, a \textit{wrapper} can be defined (Box~\ref{code:rf-wrapper}):

\begin{rcode}[label={code:rf-wrapper}]{Random Forest variant with custom wrapper}
	SL.rf.better <- function(...) {
		SL.randomForest(..., ntree = 3000)
	}
	set.seed(123)
	cv_sl_rf <- CV.SuperLearner(
	Y = Y_train, X = X_train,
	family = binomial(),
	V = 5,
	SL.library = c(
	"SL.mean",
	"SL.glmnet",
	"SL.rf.better",
	"SL.randomForest"
	)
	)
	summary(cv_sl_rf)
\end{rcode}

\routputcaption{box:rf-trees}{Cross-validated risk after including two Random Forest configurations.}
\begin{routputboxref}[label={box:rf-trees}]
	\ttfamily
	\centering
	\begin{tabular}{lrrrr}
		Algorithm & Ave & se & Min & Max \\
		Super Learner & 0.16 & 0.0115 & 0.13 & 0.21 \\
		Discrete SL & 0.16 & 0.0117 & 0.13 & 0.22 \\
		SL.mean\_All & 0.25 & 0.0031 & 0.24 & 0.27 \\
		SL.glmnet\_All & 0.16 & 0.0117 & 0.13 & 0.22 \\
		SL.rf.better\_All & 0.17 & 0.0119 & 0.15 & 0.22 \\
		SL.randomForest\_All & 0.17 & 0.0120 & 0.15 & 0.22 \\
	\end{tabular}
\end{routputboxref}

The Random Forest variant with 3000 trees achieves an average cross-validated risk of $0.16742$, compared with $0.16872$ for the default configuration. Although the larger forest produces a slightly lower risk in this example, the difference is small. This comparison illustrates that increasing the number of trees does not necessarily lead to a substantial improvement in predictive performance. By including both configurations in the library, SL can evaluate them using the same cross-validation procedure rather than requiring a single configuration to be selected in advance.

When only a single configuration needs to be modified, a \textit{wrapper} is sufficient. To generate variants automatically, \texttt{create.Learner()} can be used (Box~\ref{code:create-learner}):

\begin{rcode}[label={code:create-learner}]{Generating a variant with create.Learner}
	learners <- create.Learner(
	"SL.randomForest",
	params = list(ntree = 3000)
	)
	learners$names
\end{rcode}

\routputcaption{box:create-learner-name}{Random Forest variant generated by \texttt{create.Learner()}.}
\begin{routputref}[label={box:create-learner-name}]
	[1] "SL.randomForest_1"
\end{routputref}

The function generates a new candidate that retains \texttt{SL.randomForest} while modifying the specified value of \texttt{ntree}. This approach is particularly useful when several configurations need to be evaluated.

\subsection{Hyperparameter Grids}

The evaluation can be extended to several values of a hyperparameter. In Random Forest, \texttt{mtry} determines the number of variables considered as candidates at each split.\cite{Probst2019} With the five covariates used in this example, the reference value is (Box~\ref{code:mtry-default}):

\begin{rcode}[label={code:mtry-default}]{Reference value of mtry}
	floor(sqrt(ncol(X_train)))
\end{rcode}

\routputcaption{box:mtry-default}{Reference value of \texttt{mtry} for the simulated data.}
\begin{routputref}[label={box:mtry-default}]
	[1] 2
\end{routputref}

Based on this value, three \texttt{mtry} configurations are generated (Box~\ref{code:mtry-grid}):

\begin{rcode}[label={code:mtry-grid}]{Generating mtry configurations}
	mtry_seq <- floor(
	sqrt(ncol(X_train)) * c(0.5, 1, 2)
	)
	
	learners_mtry <- create.Learner(
	"SL.randomForest",
	tune = list(mtry = mtry_seq)
	)
	
	learners_mtry$names
\end{rcode}

\routputcaption{box:mtry-learners}{Random Forest candidates generated for different values of \texttt{mtry}.}
\begin{routputref}[label={box:mtry-learners}]
	[1] "SL.randomForest_1" "SL.randomForest_2" "SL.randomForest_3"
\end{routputref}

These variants can be incorporated directly into the library (Box~\ref{code:cvsl-mtry}):

\begin{rcode}[label={code:cvsl-mtry}]{Evaluating mtry configurations with CV.SuperLearner}
	set.seed(123)
	
	cv_sl_mtry <- CV.SuperLearner(
	Y = Y_train, X = X_train,
	family = binomial(),
	V = 5,
	SL.library = c(
	"SL.mean",
	"SL.glmnet",
	learners_mtry$names,
	"SL.randomForest"
	)
	)
	summary(cv_sl_mtry)
\end{rcode}

\routputcaption{box:mtry-summary}{Cross-validated risk for the Random Forest hyperparameter configurations.}
\begin{routputboxref}[label={box:mtry-summary}]
	\ttfamily
	\centering
	\begin{tabular}{lrrrr}
		Algorithm & Ave & se & Min & Max \\
		Super Learner & 0.16 & 0.0118 & 0.13 & 0.21 \\
		Discrete SL & 0.16 & 0.0120 & 0.13 & 0.22 \\
		SL.mean\_All & 0.25 & 0.0031 & 0.24 & 0.27 \\
		SL.glmnet\_All & 0.16 & 0.0116 & 0.13 & 0.22 \\
		SL.randomForest\_1\_All & 0.16 & 0.0126 & 0.13 & 0.21 \\
		SL.randomForest\_2\_All & 0.17 & 0.0120 & 0.15 & 0.22 \\
		SL.randomForest\_3\_All & 0.17 & 0.0129 & 0.15 & 0.23 \\
		SL.randomForest\_All & 0.17 & 0.0119 & 0.15 & 0.22 \\
	\end{tabular}
\end{routputboxref}

The cross-validated results show how the choice of \texttt{mtry} affects the predictive performance of Random Forest. Among the evaluated configurations, \texttt{SL.randomForest\_1}, corresponding to \texttt{mtry = 1}, achieves the lowest average risk ($0.15868$). Increasing \texttt{mtry} to $2$ and $4$ results in average risks of $0.16855$ and $0.17494$, respectively. These results illustrate that using more candidate variables at each split does not necessarily improve predictive performance. Rather than selecting a single hyperparameter value in advance, SL allows different configurations to be included as separate candidates and evaluated within the same cross-validation procedure. \\
The same strategy can be extended to several hyperparameters simultaneously. For example, an XGBoost grid can be constructed by combining the number of trees, maximum tree depth, and learning rate (Box~\ref{code:xgb-grid}):\cite{Chen2016}

\begin{rcode}[label={code:xgb-grid}]{Generating XGBoost hyperparameters grid}
	tune_xgb <- list(
	ntrees    = c(200, 500, 1000, 2000, 5000),
	max_depth = 1:4,
	shrinkage = c(0.001, 0.01, 0.1)
	)
	
	learners_xgb <- create.Learner(
	"SL.xgboost",
	tune = tune_xgb,
	detailed_names = TRUE,
	name_prefix = "xgb"
	)
	
	cat(
	"Number of XGBoost configurations:",
	length(learners_xgb$names),
	"\n"
	)
\end{rcode}

\routputcaption{box:xgb-configurations}{Number of XGBoost hyperparameter configurations generated.}
\begin{routputref}[label={box:xgb-configurations}]
	Number of XGBoost configurations: 60
\end{routputref}

In this case, combining five values for the number of trees, four values for maximum depth, and three learning rates generates 60 candidates. The principle is the same as that used with Random Forest, although increasing the number of configurations also increases the computational cost of the evaluation.

\section{Feature Screening}

In the previous section, the SL library was expanded by incorporating different configurations of the same algorithm. Another possibility is to modify the set of covariates used by each candidate through variable-selection procedures, or \textit{feature screening}. In this case, the screening procedure is applied before fitting the algorithm, and the resulting combination is incorporated into the library as an additional candidate. As noted by Phillips et al.\cite{Phillips2023} and Williamson et al.\cite{Williamson2025}, screeners combined with learners can improve performance in high-dimensional settings and should be evaluated using CV alongside other candidates. \\
To illustrate this approach, \texttt{SL.glmnet} using all covariates is compared with a variant preceded by \texttt{screen.corP}, which selects variables according to their association with the outcome (Box~\ref{code:screen-corp}):

\begin{rcode}[label={code:screen-corp}]{Feature screening based on correlation}
	set.seed(123)
	
	cv_sl_screen <- CV.SuperLearner(
	Y = Y_train, X = X_train,
	family = binomial(),
	V = 5,
	SL.library = list(
	"SL.mean",
	"SL.glmnet",
	c("SL.glmnet", "screen.corP")))
	
	summary(cv_sl_screen)
\end{rcode}

\routputcaption{box:screen-corp}{Cross-validated risk with and without correlation-based feature screening.}
\begin{routputboxref}[label={box:screen-corp}]
	\ttfamily
	\centering
	\begin{tabular}{lrrrr}
		Algorithm & Ave & se & Min & Max \\
		Super Learner & 0.16 & 0.0118 & 0.13 & 0.22 \\
		Discrete SL & 0.16 & 0.0119 & 0.13 & 0.22 \\
		SL.mean\_All & 0.25 & 0.0031 & 0.24 & 0.27 \\
		SL.glmnet\_All & 0.16 & 0.0118 & 0.13 & 0.22 \\
		SL.glmnet\_screen.corP & 0.16 & 0.0119 & 0.13 & 0.22 \\
	\end{tabular}
\end{routputboxref}

The version of \texttt{SL.glmnet} preceded by \texttt{screen.corP} has an average risk that is nearly identical to that of \texttt{SL.glmnet} using all covariates. In this example, which contains only five predictors, there is limited scope for reducing the set of variables and, consequently, the screening step produces little change in the observed performance. In problems involving a larger number of covariates, screening procedures can become more relevant by selecting a subset of variables before fitting the algorithm. \\
The screening strategy can also be changed without modifying the subsequent learning algorithm. For example, \texttt{screen.glmnet} uses a LASSO-based procedure to determine which covariates are retained (Box~\ref{code:screen-glmnet}):

\begin{rcode}[label={code:screen-glmnet}]{LASSO-based feature screening}
	set.seed(123)
	
	cv_sl_screen2 <- CV.SuperLearner(
	Y = Y_train, X = X_train,
	family = binomial(),
	V = 5,
	SL.library = list(
	"SL.mean",
	"SL.glmnet",
	c("SL.glmnet", "screen.glmnet")))
	
	summary(cv_sl_screen2)
\end{rcode}

\routputcaption{box:screen-glmnet}{Cross-validated risk with and without LASSO-based feature screening.}
\begin{routputboxref}[label={box:screen-glmnet}]
	\ttfamily
	\centering
	\begin{tabular}{lrrrr}
		Algorithm & Ave & se & Min & Max \\
		Super Learner & 0.16 & 0.0117 & 0.13 & 0.22 \\
		Discrete SL & 0.16 & 0.0118 & 0.13 & 0.22 \\
		SL.mean\_All & 0.25 & 0.0031 & 0.24 & 0.27 \\
		SL.glmnet\_All & 0.16 & 0.0115 & 0.13 & 0.22 \\
		SL.glmnet\_screen.glmnet & 0.16 & 0.0119 & 0.13 & 0.22 \\
	\end{tabular}
\end{routputboxref}

The results are again similar to those obtained with \texttt{SL.glmnet} without prior screening.\cite{Williamson2025} These procedures become more relevant as the number of covariates increases, since different combinations of learning algorithms and \textit{screeners} can be incorporated into the library simultaneously and evaluated as distinct candidates.

\section{Optimizing for AUC}

SL allows the meta-learner criterion to be changed; for binary outcomes, \texttt{method.AUC} uses the area under the ROC curve directly as the optimization objective during weight estimation (rather than only for evaluation), with Phillips et al.\cite{Phillips2023} recommending AUC for balanced outcomes (prevalence ~50\%) and AUC-PR for imbalanced ones. In the example (Box~\ref{code:cvsl-auc}), \texttt{CV.SuperLearner} with \texttt{method.AUC} yields cross-validated AUCs of $0.82$-$0.83$ for SL, \texttt{SL.glmnet}, \texttt{SL.randomForest}, and \texttt{SL.glmnet} with \texttt{screen.corP}, versus $0.50$ for \texttt{SL.mean} (Box~\ref{box:cvsl-auc-summary}). Other meta-learners include \texttt{method.NNloglik} (negative log-likelihood for calibrated probabilities), \texttt{method.CC\_LS} (constrained least squares for continuous outcomes), and the default \texttt{method.NNLS}; the choice depends on whether the goal is discrimination (AUC), calibration (NNloglik), or general-purpose squared-error minimization (NNLS), with \texttt{CV.SuperLearner()} used to compare them empirically.

\begin{rcode}[label={code:cvsl-auc}]{CV.SuperLearner with method.AUC}
	set.seed(123)
	cv_sl_auc <- CV.SuperLearner(
	Y = Y_train, X = X_train,
	family = binomial(),
	V = 5,
	method = "method.AUC",
	SL.library = list("SL.mean", "SL.glmnet", "SL.randomForest",
	c("SL.glmnet", "screen.corP"))
	)
	summary(cv_sl_auc)
\end{rcode}

\routputcaption{box:cvsl-auc-summary}{Cross-validated AUC estimates using \texttt{method.AUC} as the meta-learner.}
\begin{routputboxref}[label={box:cvsl-auc-summary}]
	\ttfamily
	\centering
	\begin{tabular}{lrrrr}
		Algorithm & Ave & se & Min & Max \\
		Super Learner & 0.82 & NA & 0.70 & 0.87 \\
		Discrete SL & 0.82 & NA & 0.70 & 0.89 \\
		SL.mean\_All & 0.50 & NA & 0.50 & 0.50 \\
		SL.glmnet\_All & 0.83 & NA & 0.70 & 0.89 \\
		SL.randomForest\_All & 0.81 & NA & 0.70 & 0.86 \\
		SL.glmnet\_screen.corP & 0.83 & NA & 0.70 & 0.89 \\
	\end{tabular}
\end{routputboxref}

The \texttt{se} column contains \texttt{NA} values because \texttt{summary.CV.SuperLearner()} does not provide standard error estimates for \texttt{method.AUC}.\\
\noindent\textbf{Practical advice:} As the candidate library grows, external cross-validation can become computationally demanding. Parallel execution strategies for \texttt{CV.SuperLearner()} and \texttt{mcSuperLearner()} are discussed in Section~\ref{sec:parallelization}.

\section{Parallelization}\label{sec:parallelization}

The inclusion of different hyperparameter configurations and \textit{screening} procedures can increase the number of candidates in the library and, consequently, the number of model fits required during CV. Part of this computational cost can be reduced through parallelization, since the external folds of \texttt{CV.SuperLearner()} can be processed independently across multiple cores.

\subsection{Multicore Execution}

On a single machine, parallel execution can be enabled using the argument \texttt{parallel = "multicore"}. First, the number of available cores is identified and a random-number generator suitable for parallel processing is specified (Box~\ref{code:parallel-setup}):\cite{LEcuyer2002}

\begin{rcode}[label={code:parallel-setup}]{Configuring parallel execution}
	num_cores <- RhpcBLASctl::get_num_cores()
	cat("Cores available on this machine:", num_cores, "\n")
	
	options(mc.cores = num_cores)
	set.seed(123, "L'Ecuyer-CMRG")
\end{rcode}

\routputcaption{box:parallel-cores}{Number of processing cores available for parallel execution.}
\begin{routputref}[label={box:parallel-cores}]
	Cores available on this machine: 4
\end{routputref}

The library and CV procedure remain unchanged; only the execution of the external folds is distributed across multiple cores (Box~\ref{code:cvsl-parallel}):

\begin{rcode}[label={code:cvsl-parallel}]{Parallel execution of CV.SuperLearner}
	system.time({
		cv_sl_par <- CV.SuperLearner(
		Y = Y_train, X = X_train,
		family = binomial(),
		V = 5,
		parallel = "multicore",
		SL.library = c(
		"SL.mean",
		"SL.glmnet",
		"SL.randomForest"))
	})
\end{rcode}

\routputcaption{box:parallel-time}{Computational time for parallel \texttt{CV.SuperLearner()}.}
\begin{routputref}[label={box:parallel-time}]
	user  system elapsed
	16.150   2.059  13.090
\end{routputref}

In the previously presented sequential execution, the elapsed time was $20.128$ seconds, whereas the parallel execution required $13.090$ seconds. This reduction results from distributing the workload across multiple cores and can become increasingly useful as the number of candidates grows. Execution times depend on the available hardware and the library's complexity, so this comparison illustrates the computational benefit observed in this example. \\
The resulting object can be examined using \texttt{summary(cv\_sl\_par)} in the same way as in the sequential execution. Parallelization changes how the computations are distributed, but it does not change the interpretation of the risk estimates or the candidate algorithms being evaluated.

\subsection{\texttt{mcSuperLearner()}}

When the objective is to fit the final ensemble for prediction rather than to perform an external CV assessment, \texttt{mcSuperLearner()} can be used (Box~\ref{code:mcsuperlearner}). This function parallelizes the internal CV used to estimate the ensemble weights, but does not add the external evaluation layer implemented by \texttt{CV.SuperLearner()}.

\begin{rcode}[label={code:mcsuperlearner}]{Parallelized SL fit}
	set.seed(123, "L'Ecuyer-CMRG")
	
	sl_mc <- mcSuperLearner(
	Y = Y_train, X = X_train,
	family = binomial(),
	SL.library = c(
	"SL.mean",
	"SL.glmnet",
	"SL.randomForest"))
	
	sl_mc
\end{rcode}

\routputcaption{box:mc-superlearner}{Parallel SL fit using \texttt{mcSuperLearner()}.}
\begin{routputboxref}[label={box:mc-superlearner}]
	\ttfamily
	\centering
	\begin{tabular}{lrr}
		& Risk & Coef \\
		SL.mean\_All         & 0.2485 & 0.0000 \\
		SL.glmnet\_All       & 0.1573 & 0.8411 \\
		SL.randomForest\_All & 0.1699 & 0.1589 \\
	\end{tabular}
\end{routputboxref}

The distinction between the two functions therefore depends on the objective of the analysis. \texttt{CV.SuperLearner()} incorporates an external CV procedure to evaluate performance. By contrast, \texttt{mcSuperLearner()} provides a parallel fit of the ensemble that can subsequently be used to generate predictions. \\
Parallelization can also be extended to multiple processes or machines through a cluster. In this setting, worker processes are created in advance, the required data are made available to them, and the cluster object is supplied through the \texttt{parallel} argument (Box~\ref{code:cluster-parallel}):

\begin{rcode}[label={code:cluster-parallel}]{Parallel execution using a cluster object}
	cl <- parallel::makeCluster(4)
	
	# Load required packages in each worker
	parallel::clusterEvalQ(cl, {
		library(SuperLearner)
		library(glmnet)
		library(randomForest)
	})
	set.seed(123)
	cv_sl_snow <- CV.SuperLearner(
	Y = Y_train, X = X_train,
	family = binomial(),
	V = 10,
	parallel = cl,
	SL.library = c(
	"SL.mean",
	"SL.glmnet",
	"SL.randomForest"))
	parallel::stopCluster(cl)
\end{rcode}

This alternative preserves the same logic of \texttt{CV.SuperLearner()}, while allowing the computation to be distributed beyond the cores available on a single machine.

\section{Custom Wrappers}

The SL library can be extended beyond the learners and \textit{screening} procedures available in the package. To incorporate an external algorithm, it must be adapted through a \textit{wrapper} that follows the interface expected by SL. In the case of a learner, the function receives the training data and new data, fits the algorithm, and returns the predictions (\texttt{pred}) together with the fitted object (\texttt{fit}).

\subsection{Custom Learners}

As a first example, a wrapper is defined for Gradient Boosting Machine (GBM) (Box~\ref{code:gbm-wrapper}).\cite{Friedman2001Boost} The function receives \texttt{Y}, \texttt{X}, \texttt{newX}, \texttt{family}, and the observation weights. The distribution used by GBM is determined according to the outcome type, and the number of iterations is selected through the internal CV implemented in \texttt{gbm.perf()}.

\begin{rcode}[label={code:gbm-wrapper}]{Custom wrapper for GBM}
	SL.gbm2 <- function(Y, X, newX, family, obsWeights,
	gbm.trees = 5000, interaction.depth = 2,
	shrinkage = 0.001, ...) {
		form <- as.formula(paste("Y ~", paste(colnames(X), collapse = " + ")))
		
		distribution <- switch(family$family,
		gaussian = "gaussian",
		binomial = "bernoulli")
		
		fit.gbm <- gbm::gbm(formula = form, data = X,
		distribution = distribution,
		n.trees = gbm.trees,
		interaction.depth = interaction.depth,
		shrinkage = shrinkage,
		cv.folds = 5, keep.data = TRUE,
		weights = obsWeights, verbose = FALSE)
		
		best.iter <- gbm::gbm.perf(fit.gbm, method = "cv", plot.it = FALSE)
		pred <- predict(fit.gbm, newdata = newX, best.iter, type = "response")
		
		out <- list(pred = pred, fit = list(object = fit.gbm, n.trees = best.iter))
		class(out$fit) <- "SL.gbm"
		return(out)
	}
\end{rcode}

The \texttt{SL.gbm2} function adapts GBM fitting and prediction to the format required by SL. The name distinguishes this wrapper from \texttt{SL.gbm}, which is already available in the package. The \texttt{pred} and \texttt{fit} objects provide the elements required for this function to be subsequently used as a candidate in the library. \\
The same structure can be applied to algorithms with different fitting mechanisms. The following wrapper incorporates Multivariate Adaptive Regression Splines (MARS) through the \texttt{earth} package (Box~\ref{code:earth-wrapper}):

\begin{rcode}[label={code:earth-wrapper}]{Custom wrapper for MARS}
	SL.earth2 <- function(Y, X, newX, family, obsWeights, id,
	degree = 2, penalty = 3,
	nk = max(21, 2 * ncol(X) + 1),
	pmethod = "backward", ...) {
		glm_arg <- NULL
		if (family$family == "binomial" && all(Y %in% c(0, 1))) {
			glm_arg <- list(family = binomial())
		}
		fit.earth <- earth::earth(x = X, y = Y, degree = degree,
		nk = nk, penalty = penalty,
		pmethod = pmethod, glm = glm_arg)
		pred <- predict(fit.earth, newdata = newX, type = "response")
		out  <- list(pred = pred, fit = list(object = fit.earth))
		class(out$fit) <- "SL.earth"
		return(out)
	}
\end{rcode}

In \texttt{SL.earth2}, the \texttt{degree}, \texttt{nk}, and \texttt{penalty} arguments control specific aspects of the MARS fit. For a binary outcome, a logistic specification is included. Although its internal fitting procedure differs from that used by GBM, the wrapper retains the same general structure for input, prediction, and return of the fitted model.

\subsection{Custom Screening}

The possibility of defining custom functions also extends to the variable-selection procedures introduced previously. In this case, the function does not return predictions, but rather a logical vector identifying the covariates that should be retained before fitting the learner. \\
The following screener uses LASSO to select variables (Box~\ref{code:custom-screener}).\cite{Tibshirani1996,Friedman2010} If the fit does not produce an adequate selection, an alternative ranking based on Cramér's V is used.

\begin{rcode}[label={code:custom-screener}]{Custom variable screening procedure}
	screen.glmnet3 <- function(Y, X, family, alpha = 1, minscreen = 2,
	nfolds = 10, nlambda = 200, ...) {
		myfamily <- if (family$family == "binomial" && !all(Y %in% c(0, 1)))
		"gaussian" else family$family
		
		if (!is.matrix(X)) X <- model.matrix(~ -1 + ., data = X)
		
		fitCV <- tryCatch(
		glmnet::cv.glmnet(x = X, y = Y, lambda = NULL,
		type.measure = "deviance", nfolds = nfolds,
		family = myfamily, alpha = alpha,
		nlambda = nlambda, keep = TRUE),
		error = function(e) NULL
		)
		
		whichVariable <- NULL
		if (!is.null(fitCV)) {
			si <- abs((max(fitCV$nzero) - ncol(X) / 2) - fitCV$nzero)
			w2 <- as.numeric(coef(fitCV$glmnet.fit, s = fitCV$lambda.min))[-1] != 0
			w3 <- as.numeric(glmnet::coef.glmnet(
			fitCV, s = fitCV$lambda[si == min(si)][1]))[-1] != 0
			if (sum(w2) > 1)      whichVariable <- w2
			else if (sum(w3) > 1 & sum(w3) < ncol(X) / 2) whichVariable <- w3
		}
		if (is.null(whichVariable)) whichVariable <- screen.cramersv(Y, X)
		whichVariable
	}
	
	# Cramer's V fallback screener (works with binary outcome)
	screen.cramersv <- function(Y, X, nscreen = 6, cts.num = 10, ...) {
		var_cont <- apply(X, 2, function(x) length(unique(x)) > cts.num)
		cvdata   <- X
		cutf <- function(x) cut(x,
		unique(quantile(x, prob = seq(0, 1, 0.2))),
		include.lowest = TRUE)
		if (sum(var_cont) > 0) cvdata[, var_cont] <- apply(cvdata[, var_cont], 2, cutf)
		cvf <- function(xx) vcd::assocstats(table(Y, xx))$cramer
		whichVariable <- order(apply(cvdata, 2, cvf), decreasing = TRUE) <= nscreen
		whichVariable
	}
\end{rcode}

The procedure first attempts to select covariates using LASSO. The \texttt{tryCatch()} function prevents a problematic fit in one of the folds from interrupting the entire procedure and, when an adequate set of variables is not obtained, \texttt{screen.cramersv()} is used as an alternative. This structure is particularly relevant within CV, where the screening procedure must be repeatedly applied to different subsets of the data.

\subsection{Incorporating Custom Wrappers}

Once the functions have been defined, custom learners can be incorporated into \texttt{SL.library} in the same way as the candidates used previously. In this example, \texttt{SL.gbm2} and \texttt{SL.earth2} are evaluated together with \texttt{SL.mean} and \texttt{SL.glmnet} (Box~\ref{code:cvsl-custom}):

\begin{rcode}[label={code:cvsl-custom}]{Evaluating custom wrappers}
	set.seed(
		123,
		kind = "Mersenne-Twister",
		normal.kind = "Inversion",
		sample.kind = "Rejection"
	)
	
	cv_sl_custom <- CV.SuperLearner(
	Y = Y_train, X = X_train,
	family = binomial(), V = 5,
	SL.library = c("SL.mean", "SL.glmnet", "SL.gbm2", "SL.earth2")
	)
	summary(cv_sl_custom)
\end{rcode}

\routputcaption{box:custom-wrappers}{Cross-validated risk after incorporating custom learner wrappers.}
\begin{routputboxref}[label={box:custom-wrappers}]
	\ttfamily
	\centering
	\begin{tabular}{lrrrr}
		Algorithm & Ave & se & Min & Max \\
		Super Learner & 0.16 & 0.0116 & 0.13 & 0.22 \\
		Discrete SL & 0.16 & 0.0116 & 0.13 & 0.22 \\
		SL.mean\_All & 0.25 & 0.0031 & 0.24 & 0.27 \\
		SL.glmnet\_All & 0.16 & 0.0117 & 0.13 & 0.22 \\
		SL.gbm2\_All & 0.16 & 0.0107 & 0.13 & 0.20 \\
		SL.earth2\_All & 0.17 & 0.0130 & 0.15 & 0.22 \\
	\end{tabular}
\end{routputboxref}

The custom wrappers appear in the evaluation alongside the learners available in the package. \texttt{SL.gbm2} achieves an average risk close to $0.16$, similar to that observed for \texttt{SL.glmnet}, whereas \texttt{SL.earth2} shows a value close to $0.17$. Once adapted to the required interface, an external algorithm can therefore be evaluated and incorporated into the ensemble under the same procedure applied to the remaining candidates.

\section{Full Epidemiological Application}

The previous sections introduced the main components involved in the specification and evaluation of SL separately. This section integrates these elements into a complete epidemiological application based on the prediction of a binary clinical outcome. The analysis includes the construction of a heterogeneous library, hyperparameter variants, cross-validated evaluation, estimation of ensemble weights, and final assessment on an independent test set.

\subsection{Simulated Clinical Cohort}

A cohort of \(n=1000\) individuals is simulated to represent a clinical setting in which the outcome of interest is the occurrence of a cardiovascular event. Nine covariates are generated, including continuous and binary characteristics commonly encountered in epidemiological studies. The outcome-generating mechanism includes main effects, interactions, and a nonlinear component, producing a setting in which the underlying relationship between the predictors and the outcome is not restricted to a simple linear structure.

\begin{rcode}[label={code:clinical-cohort}]{Simulating the clinical cohort}
	set.seed(
		2024,
		kind = "Mersenne-Twister",
		normal.kind = "Inversion",
		sample.kind = "Rejection"
	)
	n_clinical <- 1000
	
	age        <- rnorm(n_clinical, 65, 10)
	bmi        <- rnorm(n_clinical, 28, 5)
	sbp        <- rnorm(n_clinical, 140, 20)
	dbp        <- rnorm(n_clinical, 85, 12)
	cholesterol<- rnorm(n_clinical, 220, 40)
	glucose    <- rnorm(n_clinical, 110, 30)
	smoking    <- rbinom(n_clinical, 1, 0.25)
	diabetes   <- rbinom(n_clinical, 1, plogis(-3 + 0.1 * (bmi - 28) + 0.01 * (glucose - 110)))
	family_hx  <- rbinom(n_clinical, 1, 0.30)
	
	# Standardised predictors for stable estimation
	stand <- function(x) (x - mean(x)) / sd(x)
	
	logit_cvd <- -2.2 +
	0.5 * stand(age) + 0.4 * stand(bmi) + 0.3 * stand(sbp) +
	0.2 * stand(cholesterol) + 0.25 * stand(glucose) +
	0.6 * smoking + 0.5 * diabetes + 0.4 * family_hx +
	0.7 * stand(age) * smoking + 0.3 * stand(sbp) * stand(cholesterol) -
	0.2 * stand(age)^2
	
	Y_cvd <- rbinom(n_clinical, 1, plogis(logit_cvd))
	cat("CVD prevalence:", round(mean(Y_cvd), 3), "\n")
\end{rcode}

The resulting prevalence of the cardiovascular event is approximately \(14\%\). The simulated cohort is subsequently organized into a data frame containing the nine predictors and the binary outcome.
\begin{rcode}[label={code:clinical-data}]{Organizing the simulated clinical data}
	clinical_data <- data.frame(age, bmi, sbp, dbp, cholesterol, glucose,
	smoking, diabetes, family_hx, Y = Y_cvd)
\end{rcode} 
This simulated setting provides a more complex prediction problem than the examples presented previously and allows the different components of the SL procedure to be evaluated jointly.

\subsection{Training and Test Sets and Library Specification}
The cohort is divided into a training sample of 700 individuals and an independent test sample of 300 individuals. The training data are used for library construction, cross-validation, and estimation of the ensemble, whereas the test sample is retained for the final evaluation of predictive performance.

\begin{rcode}[label={code:clinical-library}]{Training-test split and candidate library specification}
	set.seed(
		42,
		kind = "Mersenne-Twister",
		normal.kind = "Inversion",
		sample.kind = "Rejection"
	)
	idx <- sample(1:n_clinical, 700)
	X_clin_train <- clinical_data[idx, 1:9]
	Y_clin_train <- clinical_data[idx, 10]
	X_clin_test  <- clinical_data[-idx, 1:9]
	Y_clin_test  <- clinical_data[-idx, 10]
	
	# Core library
	sl_library <- c(
	"SL.mean", "SL.glm", "SL.glmnet", "SL.randomForest", "SL.ranger",
	"SL.earth", "SL.xgboost", "SL.rpartPrune", "SL.nnet"
	)
	
	# Hyperparameter variants
	mtry_vals    <- floor(sqrt(ncol(X_clin_train)) * c(0.5, 1, 2))
	rf_learners  <- create.Learner("SL.randomForest", tune = list(mtry = mtry_vals))
	
	xgb_tune     <- list(ntrees = c(500, 1000),
	max_depth = 3:4,
	shrinkage = c(0.01, 0.1))
	xgb_learners <- create.Learner("SL.xgboost", tune = xgb_tune,
	detailed_names = TRUE, name_prefix = "xgb")
	
	full_library <- c(sl_library, rf_learners$names, xgb_learners$names)
	cat("Total number of candidate learners:", length(full_library), "\n")
\end{rcode}

The initial library includes algorithms with different functional forms and levels of flexibility, including generalized linear models, penalized regression, tree-based methods, boosting, splines, and neural networks.\cite{Phillips2023} Different Random Forest and XGBoost configurations are additionally incorporated as separate candidates.\cite{Wright2017,Chen2016} This strategy allows alternative algorithmic specifications to be evaluated within the same cross-validation procedure rather than selecting a single configuration in advance. The resulting library contains \(20\) candidate learners.

\subsection{Cross-Validated Evaluation of the Library}
The complete library is evaluated using \texttt{CV.SuperLearner()}. Ten-fold cross-validation is implemented with stratification according to the outcome, which helps preserve the proportion of cardiovascular events across the validation partitions.\cite{Phillips2023} This consideration is particularly relevant because the outcome prevalence is approximately \(14\%\). 
The ensemble weights are estimated using negative log-likelihood through \texttt{method.NNloglik}. Under this criterion, lower values indicate better predictive performance.

\begin{rcode}[label={code:clinical-cv}]{Cross-validated evaluation of the candidate library}
	set.seed(42, "L'Ecuyer-CMRG")
	options(mc.cores = RhpcBLASctl::get_num_cores())
	system.time({
		cv_sl_clinical <- CV.SuperLearner(
		Y = Y_clin_train, X = X_clin_train,
		family = binomial(),
		method = "method.NNloglik",
		parallel = "multicore",
		cvControl = list(V = 10, stratifyCV = TRUE),
		SL.library = full_library
		)
	})
\end{rcode}

\begin{rcode}[label={code:clinical-cv-summary}]{Summarizing cross-validated performance}
	summary(cv_sl_clinical)
\end{rcode}

\routputcaption{box:clinical-cv-summary}{Cross-validated negative log-likelihood risk for SL and the candidate algorithms.}
\begin{routputboxref}[label={box:clinical-cv-summary}]
	\newcommand{\SLcvrow}[5]{%
		\noindent
		\makebox[0.48\linewidth][l]{#1}%
		\makebox[0.13\linewidth][r]{#2}%
		\makebox[0.13\linewidth][r]{#3}%
		\makebox[0.13\linewidth][r]{#4}%
		\makebox[0.13\linewidth][r]{#5}%
		\par
	}
	\ttfamily
	\raggedright
	Call:
	CV.SuperLearner(
	Y = Y\_clin\_train,
	X = X\_clin\_train,
	family = binomial(),
	SL.library = full\_library,
	method = "method.NNloglik",
	cvControl = list(V = 10, stratifyCV = TRUE),
	parallel = "multicore"
	)
	\par\medskip
	Risk is based on: Negative Log Likelihood (-2*log(L))
	\par\medskip
	All risk estimates are based on V = 10
	\par\medskip
	\SLcvrow{Algorithm}{Ave}{se}{Min}{Max}
	\SLcvrow{Super Learner}{0.35}{NA}{0.27}{0.40}
	\SLcvrow{Discrete SL}{0.35}{NA}{0.28}{0.39}
	\SLcvrow{SL.mean\_All}{0.39}{NA}{0.38}{0.41}
	\SLcvrow{SL.glm\_All}{0.35}{NA}{0.28}{0.40}
	\SLcvrow{SL.glmnet\_All}{0.35}{NA}{0.28}{0.39}
	\SLcvrow{SL.randomForest\_All}{0.37}{NA}{0.31}{0.45}
	\SLcvrow{SL.ranger\_All}{0.37}{NA}{0.31}{0.44}
	\SLcvrow{SL.earth\_All}{0.39}{NA}{0.27}{0.47}
	\SLcvrow{SL.xgboost\_All}{0.44}{NA}{0.33}{0.55}
	\SLcvrow{SL.rpartPrune\_All}{0.39}{NA}{0.38}{0.41}
	\SLcvrow{SL.nnet\_All}{Inf}{NA}{Inf}{Inf}
	\SLcvrow{SL.randomForest\_1\_All}{0.40}{NA}{0.34}{0.49}
	\SLcvrow{SL.randomForest\_2\_All}{0.37}{NA}{0.31}{0.45}
	\SLcvrow{SL.randomForest\_3\_All}{Inf}{NA}{0.32}{Inf}
	\SLcvrow{xgb\_500\_3\_0.01\_All}{0.36}{NA}{0.27}{0.40}
	\SLcvrow{xgb\_1000\_3\_0.01\_All}{0.36}{NA}{0.28}{0.42}
	\SLcvrow{xgb\_500\_4\_0.01\_All}{0.36}{NA}{0.27}{0.41}
	\SLcvrow{xgb\_1000\_4\_0.01\_All}{0.36}{NA}{0.28}{0.42}
	\SLcvrow{xgb\_500\_3\_0.1\_All}{0.40}{NA}{0.31}{0.52}
	\SLcvrow{xgb\_1000\_3\_0.1\_All}{0.44}{NA}{0.33}{0.55}
	\SLcvrow{xgb\_500\_4\_0.1\_All}{0.40}{NA}{0.30}{0.52}
	\SLcvrow{xgb\_1000\_4\_0.1\_All}{0.44}{NA}{0.33}{0.55}
\end{routputboxref}

The cross-validated results allow the performance of the SL ensemble to be compared with the individual candidate algorithms and the Discrete SL. In this example, relatively simple candidates such as \texttt{SL.glm} and \texttt{SL.glmnet} achieve low cross-validated risks despite the inclusion of substantially more flexible algorithms. \\
The neural-network candidate, \texttt{SL.nnet}, produces an infinite negative log-likelihood risk in all ten external validation folds, while \texttt{SL.randomForest\_3} produces an infinite risk in at least one fold. Such values can arise when predicted probabilities are numerically equal to zero for observed events or equal to one for observed non-events. These results illustrate the importance of examining the cross-validated performance of individual candidates, since including more flexible algorithms does not necessarily improve predictive performance. \\
The variation in the ensemble weights across the external validation partitions can be examined using the procedure introduced previously:

\begin{rcode}[label={code:clinical-weights}]{Examining ensemble weights across CV folds}
	print(review_weights(cv_sl_clinical), digits = 3)
\end{rcode}

\routputcaption{box:clinical-weight-distribution}{Distribution of ensemble weights across the external cross-validation folds.}
\begin{routputboxref}[label={box:clinical-weight-distribution}]
	\SLbodyfont
	\raggedright
	\setlength{\parskip}{0pt}
	\newcommand{\SLweightrow}[5]{%
		\noindent
		\makebox[.43\linewidth][l]{#1}%
		\makebox[.20\linewidth][r]{#2}%
		\makebox[.123\linewidth][r]{#3}%
		\makebox[.123\linewidth][r]{#4}%
		\makebox[.123\linewidth][r]{#5}%
		\par
	}
	\SLweightrow{Algorithm}{mean(weight)}{sd}{min}{max}
	\SLweightrow{SL.glm\_All}{0.51005}{0.2016}{0}{0.7496}
	\SLweightrow{SL.glmnet\_All}{0.10884}{0.2563}{0}{0.7865}
	\SLweightrow{xgb\_500\_3\_0.01\_All}{0.07316}{0.1009}{0}{0.3038}
	\SLweightrow{xgb\_500\_3\_0.1\_All}{0.05796}{0.1332}{0}{0.4124}
	\SLweightrow{SL.earth\_All}{0.05048}{0.0633}{0}{0.1862}
	\SLweightrow{SL.randomForest\_3\_All}{0.03993}{0.0708}{0}{0.2154}
	\SLweightrow{xgb\_1000\_4\_0.01\_All}{0.03912}{0.0841}{0}{0.2303}
	\SLweightrow{xgb\_500\_4\_0.1\_All}{0.03752}{0.0606}{0}{0.1329}
	\SLweightrow{SL.rpartPrune\_All}{0.01645}{0.0321}{0}{0.1023}
	\SLweightrow{SL.nnet\_All}{0.01578}{0.0160}{0}{0.0427}
	\SLweightrow{SL.xgboost\_All}{0.01326}{0.0289}{0}{0.0932}
	\SLweightrow{xgb\_1000\_4\_0.1\_All}{0.01326}{0.0289}{0}{0.0932}
	\SLweightrow{xgb\_1000\_3\_0.01\_All}{0.00963}{0.0305}{0}{0.0963}
	\SLweightrow{SL.mean\_All}{0.00622}{0.0111}{0}{0.0324}
	\SLweightrow{SL.ranger\_All}{0.00431}{0.0136}{0}{0.0431}
	\SLweightrow{SL.randomForest\_1\_All}{0.00404}{0.0128}{0}{0.0404}
	\SLweightrow{SL.randomForest\_All}{0.00000}{0.0000}{0}{0.0000}
	\SLweightrow{SL.randomForest\_2\_All}{0.00000}{0.0000}{0}{0.0000}
	\SLweightrow{xgb\_500\_4\_0.01\_All}{0.00000}{0.0000}{0}{0.0000}
	\SLweightrow{xgb\_1000\_3\_0.1\_All}{0.00000}{0.0000}{0}{0.0000}
\end{routputboxref}

In this application, \texttt{SL.glm} receives the largest average ensemble weight, while the remaining weight is distributed among several candidates. This result illustrates an important property of SL: the inclusion of highly flexible algorithms does not imply that they will dominate the final ensemble. Their contribution depends on their cross-validated predictive performance.

\subsection{Out-of-Sample Evaluation}

After cross-validated evaluation, a final SL model is fitted using the complete training sample and the same library and meta-learning criterion.\cite{vanDerLaan2007} Predictions are then generated for the 300 individuals in the independent test sample, which were not used during model construction.

\begin{rcode}[label={code:clinical-evaluation}]{Evaluating the final SL model on the test set}
	set.seed(
		42,
		kind = "L'Ecuyer-CMRG",
		normal.kind = "Inversion",
		sample.kind = "Rejection"
	)
	
	sl_clinical_full <- SuperLearner(
	Y = Y_clin_train, X = X_clin_train,
	family = binomial(), method = "method.NNloglik",
	SL.library = full_library
	)
	
	pred_clinical <- predict(sl_clinical_full, X_clin_test, onlySL = TRUE)
	
	roc_test <- pROC::roc(Y_clin_test, pred_clinical$pred, quiet = TRUE)
	cat("Test AUC:", round(as.numeric(pROC::auc(roc_test)), 4), "\n")
	cat("95% CI :", round(as.numeric(pROC::ci.auc(roc_test)), 4), "\n")
\end{rcode}

The resulting area under the receiver operating characteristic curve is

\[
\mathrm{AUC}=0.7164,
\]

with a \(95\%\) confidence interval of

\[
[0.6338,\;0.7990].
\]

These estimates describe the discriminatory ability of the fitted ensemble when applied to observations not used during model development.

\begin{rcode}[label={code:clinical-roc}]{Plotting the ROC curve on the independent test set}
	plot(roc_test,
	main = "SL on the 30% held-out test set",
	print.auc = TRUE, print.auc.y = 0.4,
	col = "#2ECC71", lwd = 3)
	abline(a = 0, b = 1, lty = 2, col = "gray50")
\end{rcode}

\begin{figure}
	\centering
	\includegraphics[width=\linewidth]{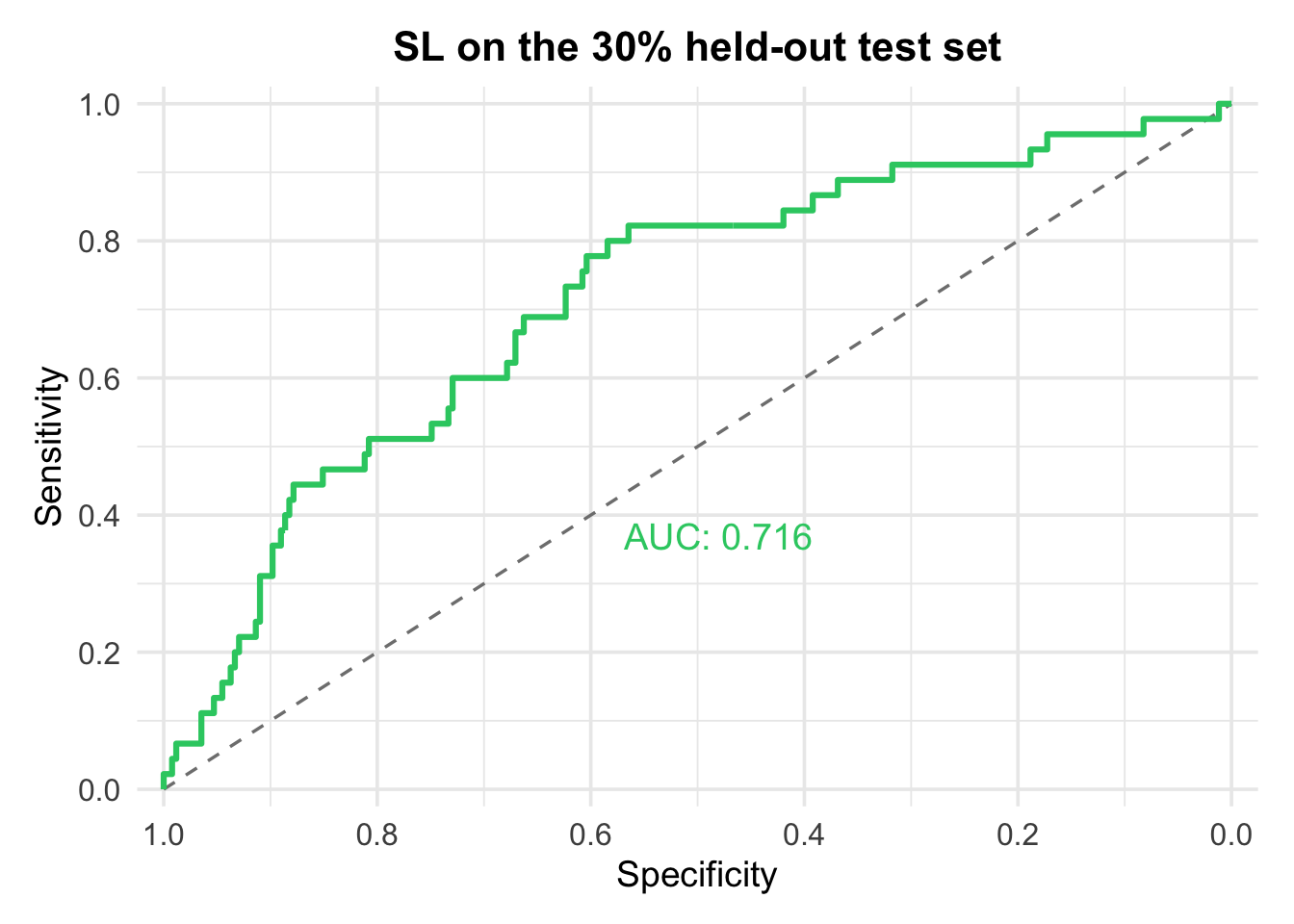}
	\caption{ROC curve for the SL predictions on the independent test set.}
	\label{fig:plt_13}
\end{figure}

Predictive performance is additionally examined through calibration. Whereas the AUC evaluates the ability of the model to discriminate between individuals with and without the outcome, calibration evaluates the agreement between predicted probabilities and observed event frequencies.

\begin{rcode}[label={code:clinical-calibration}]{Assessing calibration on the independent test set}
	cal_df <- data.frame(predicted = as.vector(pred_clinical$pred),
	observed  = Y_clin_test)
	
	cal_bins <- cut(cal_df$predicted, breaks = 10)
	cal_sum  <- aggregate(cbind(predicted, observed) ~ cal_bins,
	data = cal_df, FUN = mean)
	
	ggplot(cal_sum, aes(x = predicted, y = observed)) +
	geom_point(size = 3, color = "#2ECC71") +
	geom_smooth(method = "loess", se = FALSE, color = "#E74C3C") +
	geom_abline(intercept = 0, slope = 1, linetype = "dashed",
	color = "gray50") +
	coord_equal(xlim = c(0, 1), ylim = c(0, 1)) +
	labs(x = "Predicted probability", y = "Observed proportion") +
	theme_minimal(base_size = 14)
\end{rcode}

\begin{figure}
	\centering
	\includegraphics[width=\linewidth]{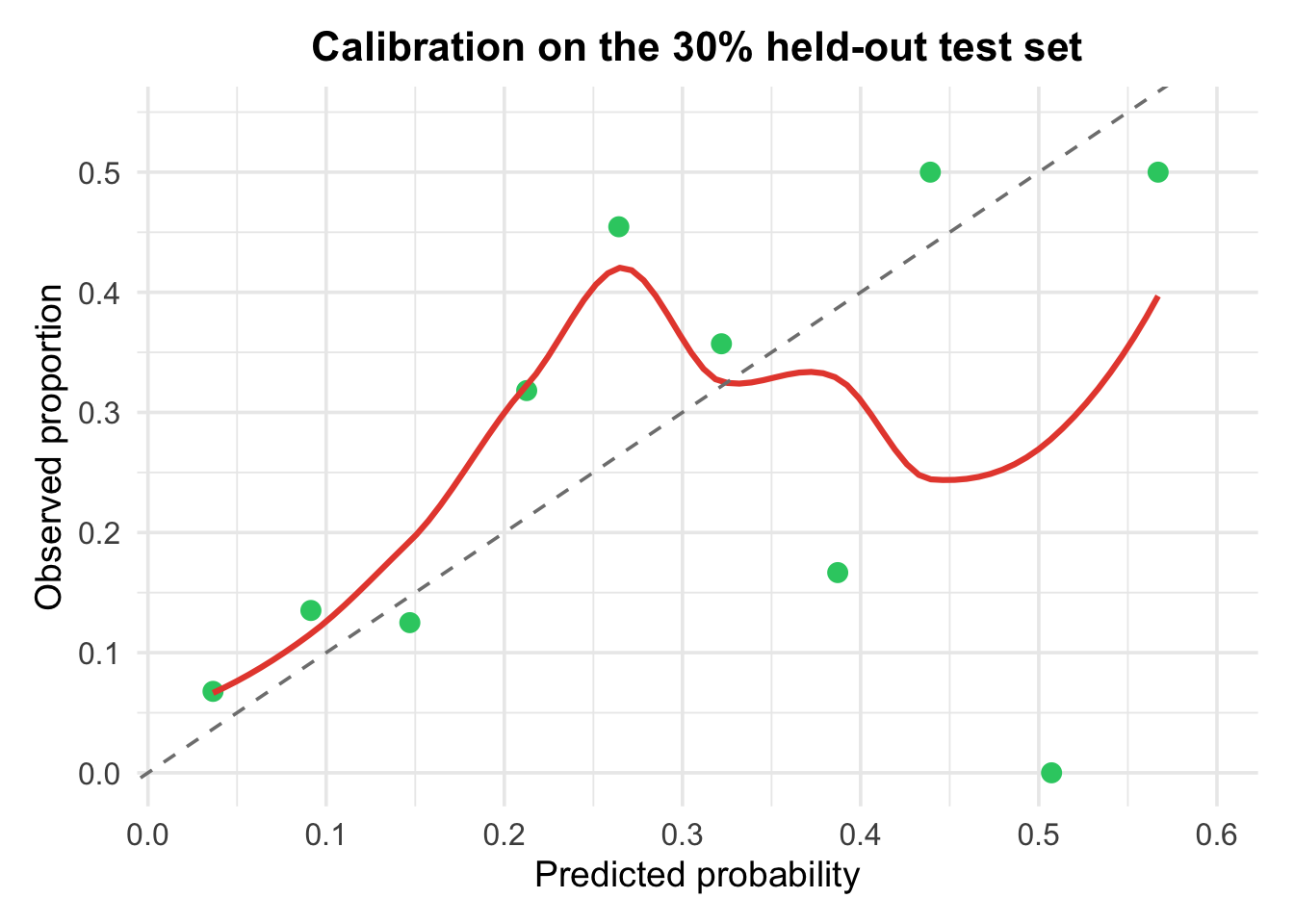}
	\caption{Calibration of SL predictions on the independent test set.}
	\label{fig:pred_13}
\end{figure}

The calibration plot compares the mean predicted probability with the observed proportion of events across groups of individuals with similar predicted risk. Agreement with the diagonal reference line indicates closer correspondence between predicted and observed risks. \\
Taken together, this application illustrates the complete SL workflow in an epidemiological prediction setting: specification of a heterogeneous candidate library, cross-validated comparison of the candidate algorithms, estimation of ensemble weights, refitting using the complete training sample, and independent evaluation of discrimination and calibration.

\section{Continuous Outcomes}

While previous sections introduced the SL framework primarily in the context of classification, the methodology extends naturally to regression.\cite{vanDerLaan2007} This section illustrates how the core procedure adapts to a continuous response, highlighting the main adjustments required in model specification and performance evaluation without repeating the foundational concepts introduced previously.

To examine this setting, a synthetic dataset with \(n=500\) observations and three continuous predictors is simulated. The outcome is generated through a data-generating mechanism containing a linear effect, a quadratic term, and a two-way interaction, thereby introducing a clear departure from a strictly linear relationship.

\begin{rcode}[label={code:continuous-data}]{Generating data with a continuous outcome}
	set.seed(
		99,
		kind = "Mersenne-Twister",
		normal.kind = "Inversion",
		sample.kind = "Rejection"
	)
	
	n_reg <- 500
	
	X_reg <- data.frame(
	x1 = rnorm(n_reg),
	x2 = rnorm(n_reg),
	x3 = rnorm(n_reg)
	)
	
	Y_reg <- with(
	X_reg,
	3 + 2 * x1 + x1^2 - 1.5 * x2 * x3 +
	rnorm(n_reg, 0, 0.5)
	)
\end{rcode}

Here, \(2x_1\) represents a linear effect, \(x_1^2\) introduces nonlinearity, and the product \(x_2x_3\) introduces an interaction between predictors. The residual error term represents variation not explained by the observed predictors.

For a continuous outcome, the SL specification is adapted by setting \texttt{family = gaussian()}. The candidate library includes algorithms with different degrees of flexibility, ranging from a simple mean-based reference learner to methods capable of representing nonlinear relationships and interactions.

\begin{rcode}[label={code:continuous-sl}]{Fitting SL with a continuous outcome}
	set.seed(
		99,
		kind = "Mersenne-Twister",
		normal.kind = "Inversion",
		sample.kind = "Rejection"
	)
	sl_reg <- SuperLearner(
	Y = Y_reg,
	X = X_reg,
	family = gaussian(),
	SL.library = c("SL.mean","SL.glm","SL.glmnet","SL.randomForest","SL.earth"))
	sl_reg
\end{rcode}

\routputcaption{box:continuous-sl}{Cross-validated risk and ensemble weights for the continuous-outcome example.}
\begin{routputboxref}[label={box:continuous-sl}]
	\ttfamily
	\centering
	\begin{tabular}{lrr}
		Algorithm & Risk & Coef \\
		SL.mean\_All         & 8.477 & 0.0000 \\
		SL.glm\_All          & 5.171 & 0.0000 \\
		SL.glmnet\_All       & 5.070 & 0.0000 \\
		SL.randomForest\_All & 1.728 & 0.0000 \\
		SL.earth\_All        & 0.284 & 1.0000 \\
	\end{tabular}
\end{routputboxref}

In this example, \texttt{SL.earth} attains by far the lowest cross-validated risk among the candidate algorithms ($0.284$ versus $1.728$ for the next best candidate) and therefore receives the entire ensemble weight ($1.0000$). The remaining candidates receive a weight of zero, so the SL prediction coincides with the prediction of \texttt{SL.earth}. This outcome is possible because the candidate library is evaluated and combined on the same cross-validation folds: when one learner clearly dominates the others, the non-negative least-squares solution places all the weight on that learner. \\
For continuous outcomes, the default optimization method, \texttt{method.NNLS}, uses squared-error loss to estimate the ensemble weights. The risk values reported above therefore correspond to cross-validated squared-error risk. \\
Once the ensemble has been fitted, predictions can be generated and additional measures of prediction error can be calculated. In this example, the mean squared error (MSE) and root mean squared error (RMSE) are used.

\begin{rcode}[label={code:continuous-error}]{Evaluating prediction error}
	pred_reg <- predict(sl_reg, X_reg)
	mse <- mean((Y_reg - pred_reg$pred)^2)
	cat("MSE :", round(mse, 4), "\n")
	cat("RMSE:", round(sqrt(mse), 4), "\n")
\end{rcode}

\routputcaption{box:continuous-error}{Mean squared error and root mean squared error for the fitted SL ensemble.}
\begin{routputref}[label={box:continuous-error}]
	MSE : 0.2368
	RMSE: 0.4866
\end{routputref}

The ensemble achieves an in-sample MSE of \(0.2368\) and an RMSE of \(0.4866\). The MSE summarizes the average squared difference between the observed and predicted outcomes, whereas the RMSE expresses prediction error on the same scale as the original outcome, facilitating its interpretation. \\
Because these measures are calculated using the same observations employed to fit the ensemble, they describe in-sample predictive error rather than generalization performance.\cite{Collinse074819,VarmaSimon2006} In applied prediction settings, performance should instead be assessed using observations not involved in model fitting, such as an independent test sample or an appropriate external cross-validation procedure. \\
Overall, the core SL workflow remains largely unchanged when moving from a binary to a continuous outcome. The principal operational changes concern the specification of \texttt{family = gaussian()} and the use of squared-error loss for estimating and evaluating the ensemble.

\section{Dose-Response Estimation}\label{sec:dose-response}

A classic epidemiological problem \cite{Greenland1995DoseResponse} is estimating the shape of an exposure-outcome relationship when it is strongly non-linear: here, a piecewise curve with a discontinuity. \\
\textbf{Why and how.} The data-generating mechanism stitches a square-root branch below 2 together with a quadratic branch above 2, and \texttt{rlaplace()} adds heavier-than-normal noise. The fine grid \texttt{xl} will be used later to draw the truth and the fits on the same scale. This is the only continuous-outcome block in the tutorial, so it also demonstrates \texttt{family = gaussian()} and \texttt{method.NNLS} (the default) with a library that pairs a flexible spline (\texttt{SL.gam}) and a hinge-function learner (\texttt{SL.earth}), which are the two workhorses for non-linear continuous problems.\cite{HastieTibshirani1986,Friedman1991}

\begin{rcode}[label={code:dose-response}]{Data generation and true curve}
	set.seed(12345, kind = "Mersenne-Twister",
					normal.kind = "Inversion", sample.kind = "Rejection")
	n_dose <- 1000
	x_dose <- runif(n_dose, 0, 8)
	# True dose-response: sqrt branch below 2, quadratic branch above 2
	y_dose <- 5 + 4 * sqrt(9 * x_dose) * as.numeric(x_dose < 2) +
	as.numeric(x_dose >= 2) * (abs(x_dose - 6)^2) + rlaplace(n_dose)
	D_dose <- data.frame(x = x_dose, y = y_dose)
	# The true curve, to compare against
	xl <- seq(0, 8, 0.1)
	yl <- 5 + 4 * sqrt(9 * xl) * as.numeric(xl < 2) +
	as.numeric(xl >= 2) * (abs(xl - 6)^2)
	truth_df <- data.frame(x = xl, y = yl)
\end{rcode}

\begin{rcode}[label={code:dose-sl-fit}]{Super Learner fit for dose-response}
	set.seed(123, kind = "Mersenne-Twister",
					normal.kind = "Inversion", sample.kind = "Rejection")
	sl_dose <- SuperLearner(
	Y = D_dose$y, X = data.frame(x = D_dose$x),
	family = gaussian(),
	method = "method.NNLS",
	SL.library = c("SL.mean", "SL.glm", "SL.gam", "SL.earth")
	)
	sl_dose
\end{rcode}

\routputcaption{box:dose-sl-fit}{Cross-validated risks and ensemble weights for the dose-response example.}
\begin{routputboxref}[label={box:dose-sl-fit}]
	\ttfamily
	\raggedright
	Call:
	SuperLearner(
	Y = D\_dose\$y, X = data.frame(x = D\_dose\$x),
	family = gaussian(),
	method = "method.NNLS",
	SL.library = c("SL.mean", "SL.glm", "SL.gam", "SL.earth")
	)
	\medskip
	\centering
	\begin{tabular}{lrr}
		& Risk & Coef \\
		SL.mean\_All   & 30.58 & 0.0000 \\
		SL.glm\_All    & 14.11 & 0.0000 \\
		SL.gam\_All    & 10.24 & 0.0084 \\
		SL.earth\_All  & 2.45  & 0.9916 \\
	\end{tabular}
\end{routputboxref}

The ensemble assigns most of its weight to \texttt{SL.earth} ($\alpha = 0.9916$), with a small contribution from \texttt{SL.gam} ($\alpha = 0.0084$). The remaining candidate algorithms receive zero weight. Among the individual candidates, \texttt{SL.earth} achieves the lowest cross-validated risk ($\mathrm{MSE} = 2.45$).

\begin{rcode}[label={code:dose-predict}]{Predict on fine grid}
	# Predict along a fine grid of "doses".
	# onlySL = FALSE so library.predict keeps the genuine curve of every
	# learner (with TRUE, zero-weight learners would be returned as 0)
	pred_dose <- predict(sl_dose, data.frame(x = xl), onlySL = FALSE)
\end{rcode}

\begin{rcode}[label={code:dose-response-plot}]{Dose-response curve plot}
	# Column order matches SL.library: 1 = mean, 2 = glm, 3 = gam, 4 = earth
	dose_df <- rbind(
	data.frame(x = xl, y = truth_df$y,              Model = "Truth"),
	data.frame(x = xl, y = pred_dose$pred,          Model = "SuperLearner"),
	data.frame(x = xl, y = pred_dose$library.predict[, 2], Model = "glm"),
	data.frame(x = xl, y = pred_dose$library.predict[, 3], Model = "gam")
	)
	
	ggplot() +
	geom_point(data = D_dose, aes(x, y),
	color = "gray75", alpha = 0.3, size = 0.6) +
	geom_line(data = dose_df, aes(x, y, color = Model), size = 1.1) +
	scale_color_manual(values = c(Truth = "black", SuperLearner = "#E74C3C",
	glm = "#3498DB", gam = "#2ECC71")) +
	labs(x = "Exposure (dose)", y = "Outcome") +
	theme_minimal(base_size = 14) +
	theme(legend.position = c(0.8, 0.75))
\end{rcode}

\begin{figure}
	\centering
	\includegraphics[width=\linewidth]{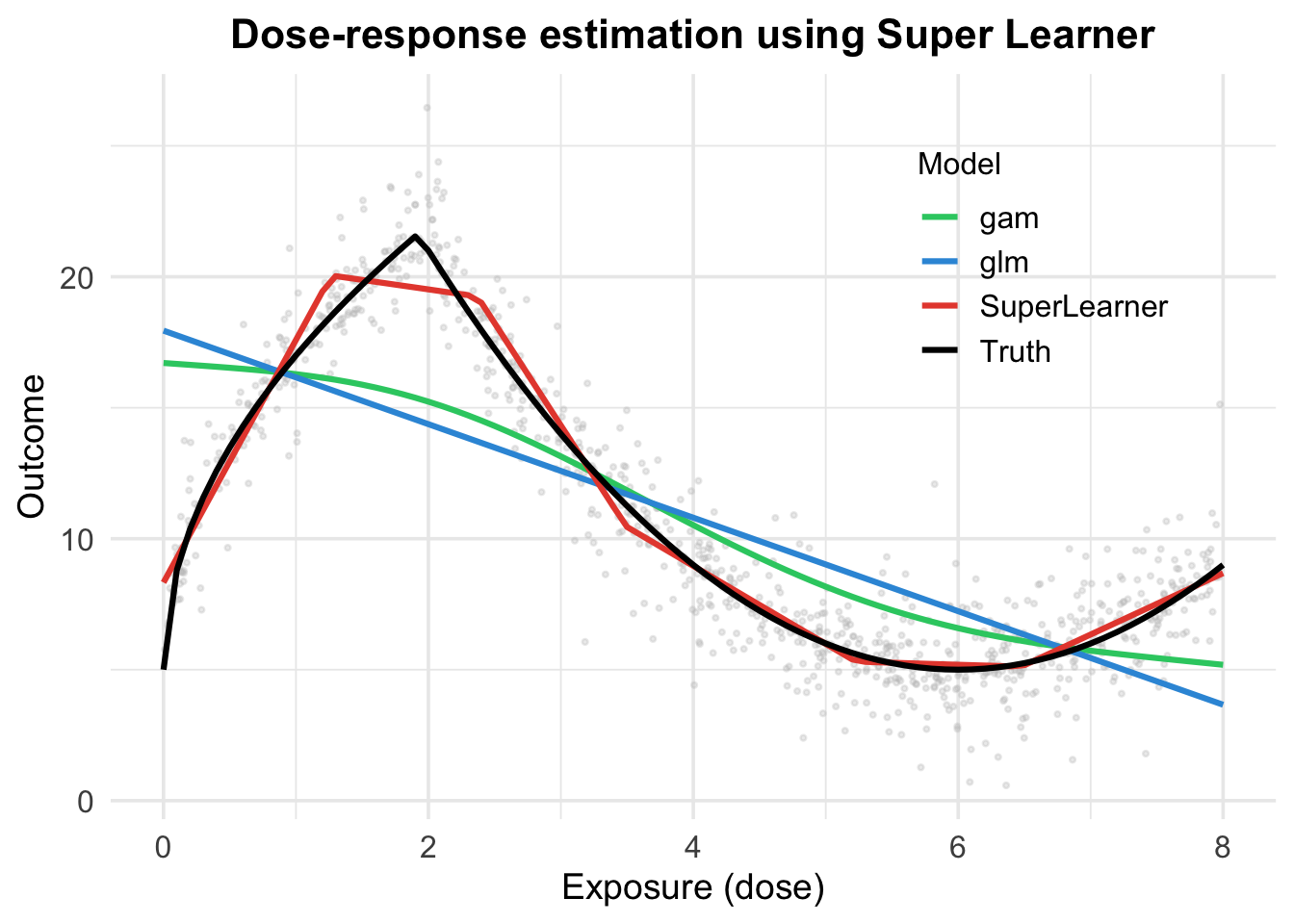}
	\caption{Dose-response curve: the Super Learner ensemble (here dominated by MARS) follows the non-linear truth more closely than the single GLM and GAM fits.}
	\label{fig:dose-response}
\end{figure}

\begin{tcolorbox}[title={Key insight}, colback=SLcodebg, colframe=SLcodeborder, arc=2mm, boxrule=0.4pt]
	A flexible learner combination (\texttt{SL.gam}, \texttt{SL.earth}) lets SL track strongly non-linear dose-response shapes without \textit{a priori} specification of the functional form. That is what makes it useful for exposure-response analyses.
\end{tcolorbox}

\section{Comparison: SL vs caret}\label{sec:sl-vs-caret}

The \texttt{caret} \cite{Kuhn2008Caret}  package provides a unified interface for hundreds of models with built-in cross-validation, preprocessing, and tuning. SL differs in philosophy and scope:

\begin{itemize}
	\item \textbf{Goal}: \texttt{caret} selects and tunes a \textit{single best model}; SL builds an \textit{ensemble} that asymptotically dominates any single candidate.
	\item \textbf{CV structure}: \texttt{caret} uses standard CV for tuning; SL uses \textit{nested} CV (internal for weights, external via \texttt{CV.SuperLearner()} for unbiased risk).
	\item \textbf{Meta-learning}: \texttt{caret} does not combine models; SL's meta-learner (NNLS, AUC, etc.) is central.
	\item \textbf{Library}: \texttt{caret} has more pre-processing options; SL has built-in screeners and a simpler wrapper interface for custom learners.
	\item \textbf{Causal inference}: SL is the foundation for TMLE and AIPW; \texttt{caret} is not designed for this.
\end{itemize}

In practice, the two approaches can be complementary: \texttt{caret} provides tools for model training, tuning, and comparison, whereas SL combines predictions from multiple candidate algorithms through cross-validation.\cite{Phillips2023,Kuhn2008Caret}

\section{Best Practices for Epidemiologists and Medical Statisticians}\label{sec:best-practices}

\subsection{Curate the library}

At a minimum include:

\begin{itemize}
	\item \textbf{\texttt{SL.glm}} -- logistic regression (the epidemiological reference);
	\item \textbf{\texttt{SL.glmnet}} -- penalised regression (high-dimensional settings);
	\item \textbf{\texttt{SL.randomForest}} / \textbf{\texttt{SL.ranger}} -- ensemble of trees;
	\item \textbf{\texttt{SL.xgboost}} -- gradient boosting;
	\item \textbf{\texttt{SL.earth}} / \textbf{\texttt{SL.gam}} -- smooth non-linear flexibility;
	\item \textbf{\texttt{SL.mean}} -- the benchmark every learner must beat.
\end{itemize}

\textbf{Why and how.} Every library you build will start from this skeleton. The rationale is simple: you want at least one linear, one penalised linear, one tree ensemble, one booster, one flexible spline, and the benchmark. The exact names are the ones that ship with SuperLearner and work on every platform; if you later discover a better learner for your domain (e.g., a specialised survival wrapper), you add it, you do not remove the basics.

\subsection{Specify the SL configuration}

Following Phillips et al.\cite{Phillips2023}, a well-specified SL requires five key decisions:

\begin{enumerate}
	\item \textbf{Loss function:} Choose the loss matching the outcome -- MSE (\texttt{method.NNLS}) for continuous outcomes; AUC (\texttt{method.AUC}) for balanced binary outcomes; AUC-PR for imbalanced binary outcomes; negative log-likelihood (\texttt{method.NNloglik}) for calibrated probabilities.\cite{Saito2015PR,VanCalster2019Calibration}
	
	\item \textbf{Effective sample size:} Compute $n_{\text{eff}}$ ($n$ for continuous $Y$; $\min(n, 5\times n_{\text{rare}})$ for binary $Y$) to guide learner complexity and number of CV folds.
	
	\item \textbf{CV scheme:} Define the $V$-fold CV scheme (typically $V=10$ or $20$; fewer folds for large $n_{\text{eff}}$, more for small; stratified for binary outcomes; cluster-level folds for dependent data via \texttt{cvControl}).
	
	\item \textbf{Diverse library:} Include parametric (GLM, penalized regression), nonparametric/flexible (random forests, boosting, GAMs, SVMs), highly adaptive (HAL), and screener-learner combinations for high-dimensional data. Include both simple and complex learners since SL downweights poor performers.
	
	\item \textbf{Discrete vs Ensemble SL:} Fit both Discrete SL (single best candidate) and Ensemble SL (convex combination via NNLS or alternatives like NNloglik/AUC), compare their cross-validated risk, and prefer eSL only if it meaningfully improves on dSL.
\end{enumerate}

These steps ensure SL is a principled, well-specified prediction tool rather than a black-box ensemble.

\subsection{Always use \texttt{CV.SuperLearner()} for reported results}

\textbf{Why and how.} Only \texttt{CV.SuperLearner()} yields a \textbf{valid} estimate of ensemble performance; \texttt{SuperLearner()} on its own does not. The difference is the outer CV layer: the former estimates what the ensemble will do on new patients; the latter estimates what its ingredients did on the training folds, which is optimistic by construction. When a paper or a protocol asks for ``the SuperLearner AUC'', you run \texttt{CV.SuperLearner()} and cite its summary.

\subsection{Explore hyperparameters for the key algorithms}

\textbf{Why and how.} A single hyperparameter grid per algorithm is enough to let the ensemble hedge its bets. The \texttt{create.Learner()} calls below generate exactly the three forest variants and eight XGBoost variants used in the clinical application earlier; uncommenting them adds no new code, just expands the library. The ensemble will down-weight the bad ones automatically, so there is no downside to a wider grid.

\begin{rcode}[label={code:best-practice-tune}]{Hyperparameter grids for key algorithms}
	rf_learners_bp <- create.Learner("SL.randomForest",
	tune = list(mtry = c(3, 5, 7)))
	xgb_learners_bp <- create.Learner("SL.xgboost", tune = list(
	ntrees = c(500, 1000), max_depth = 3:4, shrinkage = c(0.01, 0.1)
	))
\end{rcode}

\subsection{Screen when $p$ is large}

\textbf{Why and how.} The \texttt{list()} syntax pairs learners with screeners, and every pair becomes a separate library member.\cite{Williamson2025} The three-row template below is a complete starting library for a wide genomic or EHR data set: an unscreened LASSO, a correlation-screened LASSO, and a correlation-rank-screened forest. Expand the ranks as the dimension grows.

\begin{rcode}[label={code:best-practice-screen}]{Feature screening with learner-screener pairs}
	SL.library_bp <- list(
	"SL.glmnet",
	c("SL.glmnet", "screen.corP"),
	c("SL.randomForest", "screen.corRank")
	)
\end{rcode}

\subsection{Enable parallel computing}

\textbf{Why and how.} Two lines enable all cores on a single machine: \texttt{options(mc.cores = ...)} reads the core count once, and \texttt{set.seed(1, "L'Ecuyer-CMRG")} installs a parallel-safe random stream. Put these at the top of every script that calls \texttt{CV.SuperLearner()} or \texttt{mcSuperLearner()} with \texttt{parallel = "multicore"}.

\begin{rcode}[label={code:best-practice-parallel}]{Parallel computing setup}
	options(mc.cores = RhpcBLASctl::get_num_cores())
	set.seed(1, "L'Ecuyer-CMRG")   # reproducibility across cores
\end{rcode}

\subsection{Bridge to causal inference with TMLE}

\textbf{Why and how.} The TMLE workflow takes two SuperLearner libraries: one for the outcome regression $Q$ and one for the propensity score $g$. The pattern below is the recommended starting point: a rich library for $Q$ and a simpler, well-behaved one for $g$ (propensity scores are delicate and benefit from fewer, more stable learners).\cite{vanDerLaanRubin2006,Gruber2012TMLE} Uncomment and supply your own $Y$, $A$, $W$ to run it.

\begin{rcode}[label={code:tmle-connection}]{TMLE with SuperLearner}
	library(tmle)
	
	tmle_fit <- tmle(
	Y = Y, A = A, W = W,
	Q.SL.library = c("SL.glm", "SL.glmnet", "SL.randomForest", "SL.xgboost"),
	g.SL.library = c("SL.glm", "SL.glmnet"),
	V.Q = 5,
	V.g = 5
	)
\end{rcode}

\begin{tcolorbox}[title={Key insight}, colback=SLcodebg, colframe=SLcodeborder, arc=2mm, boxrule=0.4pt]
	Super Learner (or a library of $\ge$ 15 flexible algorithms) is the \textit{recommended} estimator for the nuisance parameters of TMLE and double-robust methods. It keeps the model free of the ``unrealistic modeling assumptions'' that modern causal inference is trying to avoid.
\end{tcolorbox}

\section{SL package summary Table}\label{sec:summary-table}

The following table collects the key functions, arguments, and workflow steps for quick reference.

\begin{table}[!htbp]
	\centering
	\caption{Super Learner quick-reference summary.}
	\label{tab:summary}
	\footnotesize
	\setlength{\tabcolsep}{3pt}
	\renewcommand{\arraystretch}{1.05}
	\begin{tabularx}{\linewidth}{@{}>{\raggedright\arraybackslash}p{0.30\linewidth}>{\raggedright\arraybackslash}X@{}}
		\hline
		\textbf{Task} & \textbf{Function / Argument} \\
		\hline
		Fit final ensemble & \texttt{SuperLearner(Y, X, family, SL.library)} \\
		Parallel final fit & \texttt{mcSuperLearner(Y, X, family, SL.library)} \\
		External CV evaluation & \texttt{CV.SuperLearner(Y, X, family, V, SL.library)} \\
		Predict on new data & \texttt{predict(sl\_obj, newdata, onlySL=FALSE)} \\
		Manual weighted sum & \texttt{pred\$library.predict \%*\% sl\_obj\$coef} \\
		Custom learner & Write wrapper with \texttt{pred}, \texttt{fit} return list \\
		Custom screener & Write function returning logical vector of retained variables \\
		Meta-learner (MSE) & \texttt{method.NNLS} (default) \\
		Meta-learner (AUC) & \texttt{method.AUC} (binary outcomes) \\
		Meta-learner (Log-lik) & \texttt{method.NNloglik} (binary, calibrated probs) \\
		Hyperparameter grid & \texttt{create.Learner("SL.xgboost", tune=list(...))} \\
		Screen + learner & \texttt{c("SL.glmnet", "screen.corP")} in library \\
		Extract SL weights & \texttt{sl\_obj\$coef} \\
		Extract CV risk & \texttt{cv\_sl\_obj\$cvRisk} \\
		Discrete SL winner & \texttt{cv\_sl\_obj\$whichDiscreteSL} \\
		Parallel external CV & \texttt{CV.SuperLearner(..., parallel="multicore")} \\
		Session info & \texttt{sessionInfo()} \\
		\hline
	\end{tabularx}
\end{table}

\section{SL Quick Reference Cheat Sheet}\label{sec:cheat-sheet}

\noindent\textbf{Install once:}
\begin{lstlisting}[style=SLcheat]
	install.packages(c(
	"SuperLearner", "glmnet", "randomForest",
	"ranger", "xgboost", "earth", "gam",
	"gbm", "nnls", "ggplot2", "pROC"
	))
\end{lstlisting}

\noindent\textbf{Load every session:}
\begin{lstlisting}[style=SLcheat]
	library(SuperLearner)
	library(glmnet)
	library(randomForest)
	library(ranger)
	library(xgboost)
	library(earth)
	library(gam)
	library(gbm)
	library(ggplot2)
	library(pROC)
\end{lstlisting}

\noindent\textbf{Basic SL fit (binary outcome):}
\begin{lstlisting}[style=SLcheat]
sl <- SuperLearner(
	Y = Y_train,
	X = X_train,
	family = binomial(),
	SL.library = c("SL.mean", "SL.glmnet", "SL.randomForest")
)
\end{lstlisting}

\noindent\textbf{External CV (report this):}
\begin{lstlisting}[style=SLcheat]
cv_sl <- CV.SuperLearner(
	Y = Y_train,
	X = X_train,
	family = binomial(),
	V = 10,
	SL.library = c(
		"SL.mean",
		"SL.glmnet",
		"SL.randomForest"
	)
)
summary(cv_sl)
\end{lstlisting}

\noindent\textbf{Predict on test set:}
\begin{lstlisting}[style=SLcheat]
pred <- predict(
	sl,
	newdata = X_test,
	onlySL = FALSE
)
# pred$pred = SL predictions
# pred$library.predict = individual learner predictions
\end{lstlisting}

\noindent\textbf{Add hyperparameters:}
\begin{lstlisting}[style=SLcheat]
rf_tuned <- create.Learner(
	"SL.randomForest",
	tune = list(
	mtry = c(2, 4, 6),
	ntree = c(500, 1000)
	)
)
sl <- SuperLearner(
	Y, X,
	family = binomial(),
	SL.library = c(
		"SL.mean",
		"SL.glmnet",
		rf_tuned$names
	)
)
\end{lstlisting}

\noindent\textbf{Add screening:}
\begin{lstlisting}[style=SLcheat]
sl <- SuperLearner(
	Y, X,
	family = binomial(),
	SL.library = list(
		"SL.mean",
		"SL.glmnet",
		c("SL.glmnet", "screen.corP"),
		c("SL.xgboost", "screen.glmnet")
	)
)
\end{lstlisting}

\noindent\textbf{Parallel CV:}
\begin{lstlisting}[style=SLcheat]
options(
	mc.cores = parallel::detectCores()
)
set.seed(123, "L'Ecuyer-CMRG")
cv_sl <- CV.SuperLearner(
	Y, X,
	family = binomial(),
	V = 10,
	parallel = "multicore",
	SL.library = lib
)
\end{lstlisting}

\noindent\textbf{Custom learner template:}
\begin{lstlisting}[style=SLcheat]
SL.mywrapper <- function(
Y, X, newX, family, obsWeights, ...
) {
		fit <- my_algo(Y, X, ...)
		pred <- predict(
		fit, newX,
		type = "response"
	)
		list( pred = pred, fit = list(object = fit))
}
\end{lstlisting}

\section{Use of Artificial Intelligence (AI) and AI-Assisted Technologies}

During the preparation of this work, the authors utilized Anthropic’s \textbf{Claude Code}, \textbf{Claude 3.5 Sonnet}, and \textbf{Claude 3.5 Haiku} to assist in refining English grammar, prose, and readability, as well as to review, test, and validate computer code and numerical results. The underlying methodology, initial manuscript draft, and original code were conceived and authored entirely by the human authors. All AI-assisted edits, code suggestions, and outputs were thoroughly audited, tested, and approved by the authors, who accept full responsibility for the accuracy, content, and integrity of the final publication. 

\section*{Conflicts of Interest}
The authors declare no conflicts of interest.

\bibliographystyle{unsrt}
\bibliography{references}
\end{document}